\documentclass[letterpaper,11pt]{article}
\usepackage[letterpaper,margin=2.45cm]{geometry}
\usepackage[utf8]{inputenc}
\usepackage{amsmath,amssymb,mathtools}
\numberwithin{equation}{section}
\usepackage{booktabs}
\usepackage{graphicx}
\usepackage{xspace}
\usepackage[font=small,labelfont=bf]{caption}
\usepackage{cite}
\usepackage{commath}
\usepackage{xcolor}
\usepackage{dsfont}
\usepackage[nolist]{acronym}
\usepackage[textsize=tiny, backgroundcolor=white]{todonotes}
\definecolor{darkgreen}{RGB}{0, 103, 79}
\definecolor{pink}{rgb}{0.858, 0.188, 0.478}

\usepackage[pdftex,hidelinks,colorlinks=true,citecolor=darkgreen]{hyperref}
\hypersetup{
    pdftitle = {Non-Perturbative Moduli Stabilization with Sp(4,Z) Modular Symmetry},
    pdfauthor = {Hernandez-Segura, Liu, Nilles, Ramos-Sanchez}
}

\usepackage{cleveref}

\DeclareMathOperator{\rank}{rank}
\DeclareMathOperator{\diag}{diag}
\DeclareMathOperator{\Stab}{Stab}
\DeclareMathOperator{\Sym}{Sym}
\DeclareMathOperator{\Disc}{Disc}

\DeclareMathOperator{\dv}{div}

\newcommand{\Spg}{\ensuremath{\mathrm{Sp}(4,\mathds Z)}\xspace}
\newcommand{\Siegel}{\ensuremath{\mathcal H_2}}
\newcommand{\Hess}{\ensuremath{\mathsf H}}
\newcommand{\Amod}{\ensuremath{\mathcal A_2}}
\newcommand{\Mcurv}{\ensuremath{\mathcal M_2}}
\newcommand{\Aev}{\ensuremath{A_{\rm ev}}}
\newcommand{\Isyz}{\ensuremath{\mathcal I_{\rm syz}}}

\newcommand{\e}{\mathrm{e}}
\newcommand{\dd}{\mathrm{d}}
\newcommand{\I}{\mathrm{i}}
\newcommand{\Ich}{\ensuremath{\mathsf I_\chi}}
\newcommand{\OI}{\ensuremath{\Omega_{\rm I}}}
\newcommand{\OII}{\ensuremath{\Omega_{\rm II}}}
\newcommand{\OIII}{\ensuremath{\Omega_{\rm III}}}
\newcommand{\OIV}{\ensuremath{\Omega_{\rm IV}}}
\newcommand{\OV}{\ensuremath{\Omega_{\rm V}}}
\newcommand{\OVI}{\ensuremath{\Omega_{\rm VI}}}

\renewcommand{\thefootnote}{\alph{footnote}}

\DeclareMathOperator{\re}{Re}
\DeclareMathOperator{\im}{Im}

\newcommand{\Z}[1]{\ensuremath{\mathds{Z}_{#1}}}
\newcommand{\ZZ}{\ensuremath{\mathds{Z}}}

\newcommand{\SL}[1]{\ensuremath{\mathrm{SL}(#1)}}

\newcommand{\Id}{\mathds{1}}

\newcommand{\CP}{\ensuremath{\mathcal{CP}}\xspace}
\newcommand{\x}{\ensuremath{\times}}

\everydisplay\expandafter{\the\everydisplay\thickmuskip=11mu plus 5mu}
\makeatletter
\g@addto@macro\bfseries{\boldmath}
\makeatother

\begin{document}
% %%%%%%%ACRONYMS%%%%%%%%%%%%%%%
\begin{acronym}
         \acro{CM}{complex multiplication}
\end{acronym}

\begin{titlepage}

\vspace*{1.0cm}

\begin{center}
{\Large\textbf{\boldmath
Non-Perturbative Moduli Stabilization\\[2mm]
with $\Spg$ Modular Symmetry
}
\unboldmath}

\vspace{1cm}
\textbf{Miguel Hern\'andez--Segura$^\dagger$\footnote{\texttt{hernandez@estudiantes.fisica.unam.mx}}},
\textbf{Xiang-Gan Liu$^\dagger$\footnote{\texttt{xgliu@fisica.unam.mx}}},
\textbf{Hans Peter Nilles$^*$\footnote{\texttt{nilles@uni-bonn.de}}},\\
and
\textbf{Sa\'ul Ramos--S\'anchez$^\dagger$\footnote{\texttt{ramos@fisica.unam.mx}}}
\\[5mm]
\textit{\small $^\dagger$ Instituto de F\'isica, Universidad Nacional Aut\'onoma de M\'exico, Cd.~de M\'exico,\\ C.P.~04510, M\'exico}\\[1mm]
\textit{\small $^*$ Bethe Center for Theoretical Physics and Physikalisches Institut der Universit\"at Bonn,\\ Nussallee 12, 53115 Bonn, Germany}
\end{center}

\vspace{1cm}
\vspace*{1.0cm}

\begin{abstract}
We study the stabilization of the three complex moduli associated with \Spg modular symmetry by gaugino condensation in $\mathcal N=1$ supergravity models from heterotic string theory, where one-loop threshold corrections fix both the modular weight and the singularity structure of the superpotential. We determine, for each threshold class, the complete parametrization of the consistent superpotential relying on \Spg basic modular-invariant ratios and their syzygies, generalizing the insufficient \SL{2,\ZZ}-inspired techniques based solely on polynomials of its absolute invariants.
This structure supports a rich landscape of supersymmetric Minkowski vacua. Unlike in the \SL{2,\ZZ} case, higher-order zeros produce continuous
surfaces and curves of supersymmetric Minkowski vacua, while, among the six isolated fixed points, only those with $S_4$ and $D_{12}$ residual symmetries can stabilize all three moduli.
We also construct isolated vacua away from fixed points, some of which lie at the moduli values preferred by genus-two modular flavor fits, and examine the suppression of the geometric moduli masses near an enhanced-symmetry point.
\end{abstract}

\end{titlepage}

\newpage
\setcounter{footnote}{0}
\renewcommand{\thefootnote}{\arabic{footnote}}

%=====================================================================
\section{Introduction}
\label{sec:intro}
%=====================================================================

Moduli stabilization remains a challenge in string compactifications.
In the simplest examples, string models based on toroidal compactifications include target-space $\SL{2,\ZZ}$ modular symmetries, associated with geometric complex-structure moduli $U$ and K\"ahler moduli $T$, which control, respectively, the shape and sizes of the compact space. These complex moduli appear in the low-energy action, determining the couplings among matter states~\cite{Dixon:1990pc}, and potentially driving cosmological evolution as dynamical fields~\cite{Damour:1994zq}. Even though their presence has inspired promising modular flavor phenomenology~\cite{Feruglio:2017spp}, it is also phenomenologically problematic because light moduli can mediate unobserved fifth forces and, through coherent oscillations, lead to cosmological moduli problems such as domination of the energy density or late-time entropy production~\cite{Coughlan:1983ci,deCarlos:1993wie,Banks:1995dp}. A viable theory requires moduli to be stabilized at phenomenologically admissible values. In heterotic models, this dynamics is closely related to the gauge kinetic function and one-loop threshold corrections, whose modular properties are constrained by the details of the corresponding four-dimensional supergravity effective action~\cite{Kaplunovsky:1987rp,Dixon:1990pc}. Nonperturbative effects, such as gaugino condensation~\cite{Derendinger:1985kk,Dine:1985rz}, can then lead to a scalar potential for the moduli of the target-space $\SL{2,\ZZ}$ symmetries of the model~\cite{Font:1990nt,Nilles:1990jv,Leedom:2026nby}, which are instrumental in moduli stabilization~\cite{Krasnikov:1987jj,Casas:1990qi,Cvetic:2004xx,Gaillard:2007jr,Danos:2008pv,Papineau:2009tv,Parameswaran:2010ec,Knapp-Perez:2023nty,Gordillo-Ruiz:2025xcg}.

Gaugino condensates in models based on a target-space \SL{2,\ZZ} modular symmetry induce a superpotential that takes the modular-covariant form
\begin{equation}
\label{eq:SL2Zcondensates}
  W \propto \frac{H(T)}{\eta^2(T)}\,,
\end{equation}
where $\eta(T)$ is the Dedekind eta function and $H(T)$ is a holomorphic modular invariant, which can be expressed as a rational function of the fundamental \SL{2,\ZZ}-invariant $\mathfrak j(T)$ function. This can lead to isolated stable values of the modulus $T$, corresponding to dimension $\dim_{\mathds C}=0$ loci in moduli space, which are often located at or near the fixed points of \SL{2,\ZZ}: $\I$ and $\omega:=\e^{2\pi\I/3}$, associated with remnant cyclic modular subgroups of \SL{2,\ZZ} known as stabilizers. These fixed points are also related to zeroes of the $E_4(T)$ and $E_6(T)$ generators of \SL{2,\ZZ} modular forms: $E_4(\omega)=0=E_6(\I)$. In general, modular invariance constrains the scalar potential at these fixed points. The question whether they correspond to stable minima depends on the details of the effective theory. As a final aspect, we note that the weight-one form $\eta^2(T)$ in \Cref{eq:SL2Zcondensates} vanishes only at the $\SL{2,\ZZ}$ cusp $T\to\I\infty$, so that any $H(T)$ without poles in the interior reduces to a polynomial in $\mathfrak j(T)$.

There are particular frameworks, such as \Z2 heterotic orbifolds, where the natural target-space duality group is enhanced to the Siegel modular group $\Spg$~\cite{Mayr:1995rx,Stieberger:1998yi,Nilles:2021glx}, which is the setting considered in this work. In this case, the complex structure $U$, the K\"ahler modulus $T$, and a Wilson-line modulus $Z$ are assembled into a genus-two period matrix
\begin{equation}
 \Omega=\begin{pmatrix}T&Z\\ Z&U\end{pmatrix},
 \qquad \im\Omega>0\,,
\end{equation}
and transform under Siegel modular transformations. This provides a natural generalization of the familiar \SL{2,\ZZ} modular framework. Whereas the $\eta(T)$ function plays a central role in models based on \SL{2,\ZZ}, the analogous genus-two setting naturally involves the three Igusa cusp forms $\chi_{10}, \chi_{12}$ and $\chi_{35}$, which appear in the threshold corrections~\cite{Mayr:1995rx,Nilles:1997vk,Stieberger:1998yi} and, unlike $\eta^2(T)$, vanish along complex surfaces in the interior of the complex Siegel upper half space of degree 2. This is a general feature of genus-two modular forms:
every non-zero holomorphic Siegel modular form of
positive weight has zeros in the interior of moduli
space. Moreover, any non-constant meromorphic
weight-zero modular invariant $H(\Omega)$ must have
both zeros and poles in the interior~\cite{vanderGeer:2008}.

Moreover, the notion of fixed points of the \SL{2,\ZZ} modular group generalizes in \Spg to fixed loci of complex dimensions $\dim_{\mathds C}=0,1,2$, which include six fixed points, five complex lines and two dimension-2 complex hypersurfaces in moduli space~\cite{Gottschling:1961,Gottschling:1961un,Gottschling:1967dg}. \Spg exhibits three fundamental modular invariants, called Igusa invariants, in contrast to the one included in \SL{2,\ZZ}.
Hence, in models based on \Spg, non-perturbative superpotentials generated by gaugino condensation~\cite{Derendinger:1985kk,Dine:1985rz}, analogous to \Cref{eq:SL2Zcondensates}, should incorporate all of these ingredients. One relevant question is therefore: What is the most general and consistent modular-covariant superpotential, compatible with \Spg?  Providing a complete answer to this question is the focus of this work.

As we shall see, despite the similarities of \SL{2,\ZZ} and \Spg, the naive generalization of the superpotential as a rational function of $\chi_{10}, \chi_{12}$ and $\chi_{35}$ together with polynomials of the three fundamental Igusa invariants turns out to be insufficient once the allowed pole is fixed. \Spg provides a much more complex and richer structure, as we discuss in detail in \Cref{sec:param}. In this scenario, it is natural to ask oneself what kind of phenomenological consequences may appear. One particularly interesting surprise is the emergence of supersymmetric Minkowski vacua, produced at $\dim_{\mathds C}=2$ loci in moduli space, i.e.\ there is a continuous hypersurface of Minkowski vacua, which has no \SL{2,\ZZ} analogue. We focus primarily on these supersymmetric Minkowski vacua ($W=0$) from models with \Spg modular symmetry, as they are independent of the details of dilaton stabilization. However, we also explore vacua with $W\neq 0$ at specific fixed points, which can yield stationary points with anti-de-Sitter (AdS) or de-Sitter (dS) properties depending on the dilaton sector; the task of an exhaustive scan for AdS or dS vacua is left for future work. An additional motivation to study such models endowed with a Siegel modular group is their potential to provide interesting flavor phenomenology~\cite{Ding:2020zxw,Baur:2020yjl,Ding:2021iqp}.

Our paper is organized as follows.
\Cref{sec:siegelspace} describes the details of the Siegel group and the properties of the associated moduli space. \Cref{sec:setup} sets up the supergravity framework
and threshold-condensate relations compatible with \Spg. \Cref{sec:param} is devoted to the constructions of
the complete pole-restricted function families for $\chi_{10}$, $\chi_{12}$, and $\chi_{35}$; we show how our treatment improves the standard rational generation. \Cref{sec:criteria} studies Minkowski vacuum conditions,
analyzing divisor-supported flat directions, and classifying Hessian constraints across all six elliptic
fixed-point orbits. In \Cref{sec:vacua}, we explore numerical and analytic vacua,
near-critical configurations, and multi-condensate dilaton stabilization.
\Cref{sec:discussion} summarizes some physical implications for inflation, light moduli, and modular flavor,
stressing the scope and microscopic limitations of the construction. Finally,
in \Cref{sec:conclusions} our conclusions and outlook are provided.

%=====================================================================
\section{The Siegel group and its moduli space}
\label{sec:siegelspace}
%=====================================================================

In order to fix our notation, let us briefly discuss the properties of the complex Siegel space, the Siegel group and its fundamental domain.

The complex Siegel upper half-space of degree 2, usually denoted as $\Siegel$, is parametrized by $2\times 2$ symmetric complex matrices
with a strictly positive-definite imaginary part. Its elements are represented by the genus-two modulus matrix
\begin{equation}\label{eq:Omega-def}
\Omega = \begin{pmatrix} T&Z\\ Z&U \end{pmatrix},
\qquad \im\Omega>0\,,
\end{equation}
with $T,\,U,\,Z\,\in\mathds C$. In abstract terms, $\Omega$ can be considered the parametrization of a genus-two Riemann surface (see e.g.~\cite{Brownstein:1992,Perna:2016} for details). In the context of string theory, $T,\,U$ and $Z$ correspond to the K\"ahler modulus, the complex-structure modulus, and the Wilson-line modulus, respectively. Note that the locus $Z=0$ corresponds to a decomposition into two independent modular parameters, each governed by a copy of the standard modular group $\SL{2,\ZZ}$.

On the other hand, defining the symplectic genus-two geometric form $J=\begin{pmatrix} 0 & \Id_2 \\ -\Id_2 & 0 \end{pmatrix}$, the Siegel modular group is the group of linear transformations preserving $J$, i.e.
\begin{equation}
\Spg = \left\{ \gamma \in \mathcal{M}_{4\x4}(\ZZ) \mid  \gamma^{\top} J \gamma = J  \right\}\,.
\end{equation}
Any element $\gamma\in\Spg$ can be written as
\begin{equation}
  \gamma = \begin{pmatrix} A& B \\ C & D \end{pmatrix}
  \qquad \text{with}\qquad
  A, B, C, D\in \mathcal{M}_{2\x2}(\ZZ)\,.
\end{equation}
A convenient choice of generators for \Spg is
\begin{equation}
\label{eq:Siegelgens}
  S = \begin{pmatrix} 0 & \Id_2 \\ -\Id_2 & 0 \end{pmatrix},\quad
  T_1 = \begin{pmatrix} \Id_2 & B_1 \\ 0 & \Id_2 \end{pmatrix},\quad
  T_2 = \begin{pmatrix} \Id_2 & B_2 \\ 0 & \Id_2 \end{pmatrix},\quad
  T_3 = \begin{pmatrix} \Id_2 & B_3 \\ 0 & \Id_2 \end{pmatrix},
\end{equation}
with
\begin{equation}
  B_1 =\begin{pmatrix} 1 & 0 \\ 0 & 0 \end{pmatrix},\quad
  B_2 =\begin{pmatrix} 0 & 0 \\ 0 & 1 \end{pmatrix},\quad
  B_3 =\begin{pmatrix} 0 & 1 \\ 1 & 0 \end{pmatrix}.
\end{equation}
The action of \Spg on the modulus matrix $\Omega$ is given by
\begin{equation}
\label{eq:Sp4-action}
  \gamma\Omega = (A\Omega+B)(C\Omega+D)^{-1}\,.
\end{equation}

On the other hand, the fundamental domain $\mathcal{F}_2$ of the Siegel modular group acting over $\Siegel$, is~\cite{Siegel:1943,Gottschling:1959,Ding:2021iqp}
\begin{equation}
\mathcal{F}_2 = \left\{ \Omega\in\Siegel \,\middle|\,
\begin{aligned}
\abs{\re(T)}&\leq\frac{1}{2},\,\, \abs{\re(U)}\leq\frac{1}{2},\,\,\abs{\re(Z)}\leq\frac{1}{2}\,,\\
0\leq &~ 2\im(Z)\leq \im(T)\leq\im(U),\\
1\leq &~ \abs{T},\,1\leq \abs{U},\,1\leq \abs{T+U-2Z\pm1}\,, \\
1\leq &~ \abs{\det(\Omega+\varepsilon_i)}
\end{aligned} \right\},
\end{equation}
where $\varepsilon_i$ denotes any matrix in the list
\begin{align}
\label{eq:epsilonmatrices}
\begin{pmatrix} 0 & 0\\ 0 & 0\end{pmatrix},\quad &&
\begin{pmatrix} \pm 1 & 0\\ 0 & 0\end{pmatrix},\quad &&
\begin{pmatrix} 0 & 0\\ 0 & \pm 1\end{pmatrix},\quad &&
\begin{pmatrix} \pm 1 & 0\\ 0 & \pm 1\end{pmatrix},\nonumber\\
\begin{pmatrix} \pm 1 & 0\\ 0 & \mp 1\end{pmatrix},\quad &&
\begin{pmatrix} 0 & \pm 1\\ \pm 1 & 0\end{pmatrix},\quad &&
\begin{pmatrix} \pm 1 & \pm 1\\ \pm 1 & 0\end{pmatrix},\quad &&
\begin{pmatrix} 0 & \pm 1\\ \pm 1 & \pm 1\end{pmatrix}.
\end{align}

\begin{table}[b!]
\centering
\begin{tabular}{clll}
\toprule
$\dim_{\mathds C}$ & stabilizer & moduli $(T,U,Z)$ & remark\\
\midrule
2 & $\Z2$ & $(T,U,0)$ & $H_1$, $\chi_{10}=0$\\
2 & $\Z2$ & $(T,T,Z)$ & $H_4$, $\chi_{35}=0$\\
2 & $\Z2$ & $(T,U,1/2)$ & \Spg image of $H_4$\\
\midrule
1 & $\Z4$ & $(T,\I,0)$ & inside $H_1$\\
1 & $\Z6$ & $(T,\omega,0)$ & inside $H_1$\\
1 & $\Z2\times\Z2$ & $(T,T,0)$ & $H_1\cap H_4$\\
1 & $\Z2\times\Z2$ & $(T,T,1/2)$ & inside the image of $H_4$\\
1 & $S_3$ & $(T,T,T/2)$ & inside $H_4$\\
\midrule
0 & $\Z5$ & $(\zeta,-\zeta^{-1},\zeta+\zeta^{-2})$ & $\OI$\\
0 & $S_4$ & $(\tilde\eta,\tilde\eta,\tfrac12(\tilde\eta-1))$ & $\OII$\\
0 & $(\Z4\times\Z2)\rtimes\Z2$ & $(\I,\I,0)$ & $\OIII$, node of $\dv E_6$\\
0 & $S_3\times\Z6$ & $(\omega,\omega,0)$ & $\OIV$, node of $\dv E_4$\\
0 & $S_3\times\Z2\cong D_{12}$ & $\tfrac{\I}{\sqrt3}(2,2,1)$ & $\OV$\\
0 & $\Z{12}$ & $(\omega,\I,0)$ & $\OVI$\\
\bottomrule
\end{tabular}
\caption{The thirteen inequivalent fixed-locus classes of \Spg, displayed through
fourteen convenient representatives. It is known that the second and third $\dim_{\mathds C}=2$ loci are equivalent under \Spg; we list them both because they appear as distinct boundary components of $\mathcal F_2$. Here $\omega:=\e^{2\pi\I/3}$, $\zeta:=\e^{2\pi\I/5}$ and
$\tilde\eta:=\tfrac13(1+2\sqrt2\,\I)$.}
\label{tab:fixedall}
\end{table}

A natural question is how the concept of fixed points in \SL{2,\ZZ} generalizes in \Spg. Let us define a {\it fixed locus} $\Omega_f\in\mathcal F_2$ as the modulus matrix that is left unchanged by a specific non-trivial stabilizer group
\begin{equation}
\label{eq:Sp4Z-stabilizer}
   \left\{\gamma\in\Spg ~\Big|~ \gamma\,\Omega_f \stackrel!= \Omega_f \right\}\Big/\{\pm\Id_4\}
   \subset \Spg/\{\pm\Id_4\}\,.
\end{equation}
In \Spg, this leads to various structures~\cite{Gottschling:1961}: the fixed loci are classified into 2-dimensional surfaces, 1-dimensional curves, and 0-dimensional points, as summarized in \Cref{tab:fixedall}. In each case, the dimension counts the number of complex free parameters: a curve has one and a surface has two free moduli.
Let us briefly discuss them.

Firstly, we have the 2-dimensional fixed surfaces, whose origin is associated with the transformations $Z\mapsto-Z$ and $T\leftrightarrow U$, which are respectively given by the \Spg elements $\diag(P,P)$ and $\diag(Q,Q)$ in terms of $P:=\diag(1,-1)$ and
$Q:=\bigl(\begin{smallmatrix}0&1\\1&0\end{smallmatrix}\bigr)$. Their corresponding fixed surfaces are
\begin{itemize}
    \item The Diagonal Humbert Surface $H_1$ given by $Z = 0$, fixed under $\diag(P,P)$.
    \item The Symmetric Humbert Surface $H_4$ characterized by $T = U$, fixed under $\diag(Q,Q)$. (This surface is known to be equivalent to its $\Spg$ image at $Z=1/2$.)
\end{itemize}

Secondly, we have the five 1-dimensional fixed curves: two of them lie within $H_1$, two within $H_4$ (or its image), and one in their intersection. Precisely, by demanding invariance under particular stabilizer generators~\cite{Ding:2020zxw}, they are
\begin{itemize}
    \item within $H_1$, requiring invariance under an additional elliptic transformation of $U$ fixes it to either $\I$ or $\omega$, giving the two curves described by $Z=0$, $U=\I$ and by $Z=0$, $U=\omega$;
    \item in $H_4$ we have the curve parametrized by $T=U$, $Z=T/2$, and within the $H_4$ image we have the one parametrized by $T=U$, $Z=1/2$;
    \item in $H_1\cap H_4$, imposing simultaneous invariance under $Z\mapsto -Z$ and $T\leftrightarrow U$ yields the curve $T=U$, $Z=0$.
\end{itemize}

Third, there are six isolated fixed points, labeled here $\Omega_\mathrm{I}$--$\Omega_\mathrm{VI}$, given by the representatives
\begin{align}
\label{eq:orbits_short}
 \OI   &=\begin{pmatrix}\zeta&\zeta+\zeta^{-2}\\
         \zeta+\zeta^{-2}&-\zeta^{-1} \end{pmatrix},
 &\OII &=\begin{pmatrix}\tilde\eta&\frac12(\tilde\eta-1)\\
         \frac12(\tilde\eta-1)&\tilde\eta \end{pmatrix},\nonumber\\
 \OIII &=\diag(\I,\I)\,,& \OIV&=\diag(\omega,\omega)\,,\nonumber\\
 \OV   &=\frac{\I}{\sqrt3}\begin{pmatrix}2&1\\1&2\end{pmatrix},
 &\OVI &=\diag(\omega,\I)\,.
\end{align}
All effective stabilizers, after removing the central element $-\Id_4$ that acts trivially on $\Omega$, as given by \Cref{eq:Sp4Z-stabilizer}, are listed in the second column of \Cref{tab:fixedall}. These symmetries may be regarded as remnant effective modular symmetries of the theory at the fixed loci.

These loci appear in concrete string constructions. In the symmetric
$\Z2$ Narain orbifold of Ref.~\cite{Nilles:2021glx}, the twist requires the
continuous Wilson-line background to vanish, while leaving the torus metric and
antisymmetric tensor unconstrained. The surviving moduli are therefore $(T,U,0)$.
On this locus the period matrix is diagonal, and the modular action contains
$\SL{2,\ZZ}_T\x\SL{2,\ZZ}_U$ together with the mirror transformation
$T\leftrightarrow U$. The orbifold condition fixes $Z$, but does not dynamically
select $T$ or $U$. Thus, even this familiar construction requires a potential for
its two remaining moduli. In the genus-two description, the same locus is the
zero set of $\chi_{10}$.

Interestingly, the surface $(T,T,Z)$ also has a direct application to flavor physics.
Ref.~\cite{Ding:2020zxw} restricts genus-two modular forms to $T=U$ and uses the
level-two finite modular group $S_4\times\Z2$ to construct lepton and quark mass
models. Its triplet representations allow the three families to be grouped into
matter multiplets, with Yukawa couplings determined by Siegel modular forms.
This finite group governs the transformations on the restricted moduli space;
it is distinct from the generic $\Z2$ stabilizer of the surface. In one lepton
model, a further restriction to $Z=0$ gives the fitted value
\begin{equation}
\label{eq:flavor-motivation-fit}
   T = U = 0.01541+1.00011\I\,,
\end{equation}
near the isolated point $\OIII$. The flavor fit supplies a target value for the
modulus, but does not explain how it is selected dynamically.

Finally, let us point out that the same $T=U$ surface can be realized by a Narain orbifold with the mirror
transformation as its twist~\cite{Nilles:2021glx}. The string and flavor examples
therefore lead to a common question: can modular-covariant dynamics stabilize
the surviving moduli at values compatible with the desired couplings? A fixed
locus by itself does not answer this question as some moduli are still left free.
If this arises from an orbifold compactification,
one must stabilize the fields that survive the projection. If it is a subspace
of a theory with all three moduli, one must also test fluctuations transverse to
it. We now turn to gaugino condensation as a source of the required stabilizing potential.

%=====================================================================
\section{Supergravity framework and threshold input}
\label{sec:setup}
%=====================================================================

Our setup is based on 4-dimensional $\mathcal N=1$ supergravity, with the reduced
Planck mass set to one. In this context, the K\"ahler potential of the \Spg moduli and the dilaton $S$ is
\begin{equation}
 K=-\gamma_K\log\det(\im\Omega)+K_S(S,\bar S)\,,
 \label{eq:K}
\end{equation}
where $\gamma_K>0$ is fixed by the compactification. Further, the $F$-term scalar potential in terms of the K\"ahler derivative
$D_IW:=\partial_IW+K_IW$ is given by
\begin{equation}
\label{eq:VF-setup}
   V_F = \e^K\left(K^{I\bar J}D_IW D_{\bar{J}} \overline{W}-3|W|^2\right)\,,
\end{equation}
where the indices run over all moduli, $I,\,J\in\{S,T,U,Z\}$.
Matter fields and $D$-terms are omitted in this reduced sector, because their contribution may be considered subleading. Our task
is to determine the allowed modular dependence of $W$ and then test whether and under which conditions the resulting potential stabilizes the moduli.

%%%%%%%%%%%%%%%%%%%%%%%%%%%%%%%%%%%%%%%%%%%
\subsection{The genus-one example: \SL{2,\ZZ}-covariant superpotential}

In the case of a single complex modulus $T$, the moduli-dependent superpotential is already known. A hidden asymptotically free gauge sector has a holomorphic Wilsonian gauge kinetic function $f_a(S,T)$, whose real part determines its gauge coupling. For a pure hidden gauge sector, gaugino condensation yields the superpotential contribution
\begin{equation}
 W_{\rm gc}\propto\exp\!\left(-\frac{8\pi^2}{c_a}f_a\right)\,,
 \label{eq:gc-general}
\end{equation}
where $c_a$ is the dual Coxeter number in our normalization
\cite{Derendinger:1985kk,Dine:1985rz,Cvetic:1991qm,Parameswaran:2010ec}. A perturbative threshold correction to $f_a$
therefore becomes a non-perturbative modulus-dependent factor in $W$.

In the standard $\SL{2,\ZZ}$ example, the threshold corrections to $f_a$ include the
Dedekind eta function~\cite{Kaplunovsky:1987rp,Dixon:1990pc,Font:1990nt,Nilles:1990jv}.
If the one-loop corrected gauge kinetic function is schematically expressed as
\begin{equation}
 f_a(S,T)=S-\frac{b_a^\eta}{8\pi^2}\log\eta^2(T)+f_a^{\rm inv}(T)\,,
 \label{eq:eta-kinetic}
\end{equation}
with $b_a^\eta$ fixed by the spectrum and $f_a^{\rm inv}$ modular invariant, then \Cref{eq:gc-general} leads to
\begin{equation}
 W_{\rm gc}\propto\e^{-8\pi^2S/c_a}
 \bigl[\eta^2(T)\bigr]^{b_a^\eta/c_a}
 \exp\!\left(-\frac{8\pi^2}{c_a}f_a^{\rm inv}(T)\right)\,.
 \label{eq:eta-condensate}
\end{equation}
The coefficient and the Green--Schwarz shift of $S$ ensure that the full
superpotential has the required modular weight.

For a K\"ahler term $-\gamma_T\log(2\im T)$, one can express this result in a
modular-invariant dilaton frame as
\begin{equation}
\label{eq:SL2-general-weight}
   W_{\rm np}(S,T) = B_S(S)\,\frac{H(T)}{[\eta^2(T)]^{\gamma_T}}\,.
\end{equation}
Here $H$ has weight zero, and any constant modular phase is understood in the
transformation law. The choice $\gamma_T=1$ reproduces \Cref{eq:SL2Zcondensates};
the usual overall-volume example with $\gamma_T=3$ instead has $\eta^{-6}$.
The function $\eta^2$ is the basic weight-one factor with its Dedekind multiplier,
and has no zeros in the upper half-plane. Its uniqueness holds for that weight,
multiplier and holomorphic cusp behavior, up to normalization.\footnote{In fact, for any holomorphic form $F$ with the same weight, multiplier, and cusp behavior as $\eta^2$, the ratio $F/\eta^2$ is a weight-zero function holomorphic on the upper half-plane and at the cusp, and is therefore constant.}
With meromorphic cusp behavior, the invariant freedom is carried by $H(T)$, which is rational in
the elliptic Klein invariant $\mathfrak j(T)$ and is polynomial when no interior
poles are allowed. Thus the denominator in the standard example is built from
$\eta^2$, while the vacuum structure depends on the remaining invariant factor.

%%%%%%%%%%%%%%%%%%%%%%%%%%%%%%%%%%%%%%%%%%%
\subsection{Genus-two threshold corrections and gaugino condensates}
\label{sec:threshold-condensate}

For \Spg, the structure of the superpotential contribution due to gaugino condensation remains \Cref{eq:gc-general}.
What changes is the threshold dependence on $\Omega$. With one continuous Wilson
line, the relevant Narain subsector has signature $(2,3)$ associated with the enlarged target-space duality group $\mathrm{O}(2,3;\ZZ)\cong\Spg$~\cite{Nilles:2021glx}, and its automorphic
functions are naturally organized by genus-two Siegel modular forms~\cite{Mayr:1995rx,Nilles:1997vk,Stieberger:1998yi}.
There is now a richer set of cusp forms and interior zero loci, so the modular weight alone no longer specifies a
threshold factor.

Using the action in \Cref{eq:Sp4-action}, define the automorphy factor
\begin{equation}
 j(\gamma,\Omega):=\det(C\Omega+D)\,.
 \label{eq:automorphy}
\end{equation}
A scalar Siegel modular form of weight $k$ obeys
\begin{equation}
 F_k(\gamma\Omega)=j(\gamma,\Omega)^kF_k(\Omega)\,,
 \label{eq:Siegel-form-def}
\end{equation}
with holomorphic cusp behavior. The unique non-trivial one-dimensional character
of $\Spg$ supplies an additional sign $\nu_\chi(\gamma)=\pm1$ when present.
Since \Cref{eq:K} transforms as $K\mapsto K+\gamma_K\log|j(\gamma,\Omega)|^2$, the full superpotential must carry weight
$-\gamma_K$, up to a constant phase.

The holomorphic Wilsonian coupling is schematically
\begin{equation}
 f_a(S,\Omega)=S+\frac{1}{16\pi^2}\Delta_a^{\rm hol}(\Omega)\,.
 \label{eq:f-Siegel}
\end{equation}
Near a threshold divisor, a logarithmic contribution takes the form
\begin{equation}
\label{eq:local-log}
   \Delta_a^{\rm hol}\simeq-2b_a^\star\log\Psi(\Omega)+\text{regular}\,,
\end{equation}
with $\Psi(\Omega)$ the so-called holomorphic factor. Hence,
\begin{equation}
\label{eq:gaugino_short}
   W_{\rm gc} \propto \e^{-8\pi^2S/c_a}\,\Psi(\Omega)^{b_a^\star/c_a}\,.
\end{equation}
The exponent is fixed by the spectrum; it is not a freely chosen universal pole
order. The physical gauge coupling also contains non-holomorphic terms required
by modular invariance and anomaly cancellation~\cite{Kaplunovsky:1987rp,Dixon:1990pc}. Only the holomorphic Wilsonian function
enters \Cref{eq:gc-general}. Appendix~\ref{app:thresholds} reviews this distinction
and the automorphic product formulas for the threshold corrections.

Three cusp forms are especially relevant for the threshold structures studied
here: $\chi_{10}$, $\chi_{12}$ and $\chi_{35}$, with weights $10$, $12$ and $35$,
respectively. Together with $E_4$ and $E_6$, they generate the scalar genus-two
modular-form ring, subject to the relation for $\chi_{35}^2$
\cite{Igusa:1962,Igusa:1967}. Their definitions and zero loci are collected in
Appendix~\ref{app:forms}. Their different microscopic roles also matter:
\begin{itemize}
 \item $\chi_{10}$ vanishes to order two on the separating surface $H_1$,
 represented by $Z=0$, and occurs in thresholds associated with additional
 massless states~\cite{Mayr:1995rx,Stieberger:1998yi}.
 \item $\chi_{12}$ occurs in explicit heterotic thresholds, but is non-zero on
 $H_1$ in the interior. Its zero divisor has no equally simple perturbative
 gauge-enhancement interpretation~\cite{Mayr:1995rx,Nilles:1997vk}.
 \item $\chi_{35}$ vanishes on both $H_1$ and $H_4$ and appears in universal
 threshold combinations involving other modular forms~\cite{Nilles:1997vk}.
 A threshold consisting of an isolated power of $\chi_{35}$ is a further
 compactification-dependent assumption.
\end{itemize}
The availability of these forms does not mean that an arbitrary cusp form or
exponent is generated by every compactification. We study the three choices
separately, keeping the selected threshold divisor fixed.

%%%%%%%%%%%%%%%%%%%%%%%%%%%%%%%%%%%%%%%%%%%
\subsection{Modular weight and the residual function}

For one non-perturbative contribution, write
\begin{equation}
 W_{\rm np}(S,\Omega)=W_0(S,\Omega)H(\Omega)\,,
 \label{eq:Wfactor}
\end{equation}
where $W_0$ contains the condensate exponential, threshold factor and
Green--Schwarz contribution. If $\Psi(\Omega)$ has weight $k$ and the dilaton shifts as
$S\mapsto S+\delta_{\rm GS}\log j(\gamma,\Omega)$, \Cref{eq:local-log,eq:gaugino_short}
give the condensate weight
\begin{equation}
\label{eq:weightmatch}
   w_W = -\frac{8\pi^2}{c_a}\delta_{\rm GS}+k\frac{b_a^\star}{c_a}\,.
\end{equation}
The microscopic factors must combine to reproduce $w_W=-\gamma_K$. In the
modular-invariant dilaton frame used in the main analysis, $W_0$ carries this
weight and the remaining function has weight zero:
\begin{equation}
\label{eq:Hinv}
   H(\gamma\Omega) = H(\Omega)\,,
\end{equation}
or, when the extra character is allowed,
\begin{equation}
\label{eq:Hsign}
   H(\gamma\Omega) = \nu_\chi(\gamma)H(\Omega)\,.
\end{equation}
This is the genus-two counterpart of the invariant freedom in \Cref{eq:SL2-general-weight}.

For the selected threshold, we write schematically
\begin{equation}
 W_0=\mathcal A\e^{-aS}\chi(\Omega)^{-n}\,,
 \qquad \chi\in\{\chi_{10},\chi_{12},\chi_{35}\}\,,
 \label{eq:W0schematic}
\end{equation}
where $\mathcal A$ is the condensate normalization, $a$ the dilaton exponent and
$n$ the threshold power fixed by the model. In this expression all factors needed
for the total weight are understood. Additional denominators in $H$ are restricted
to the same zero locus, so they introduce no new pole divisor.

To keep track of regularity, we distinguish the zero locus of $F$
\begin{equation}
 \mathcal Z(F):=\{\Omega\in\Siegel\mid F(\Omega)=0\}
 \label{eq:zero-locus-def}
\end{equation}
from its divisor
\begin{equation}
 \dv(F)=\sum_\alpha
 \operatorname{ord}_{\mathcal D_\alpha}(F)\mathcal D_\alpha\,,
 \label{eq:divisor-def}
\end{equation}
which is a formal linear combination of the irreducible complex surfaces (divisor components) $\mathcal D_\alpha\subset\Siegel$ on which $F(\Omega)$ vanishes or diverges, weighted by its integer vanishing (or pole) order $\operatorname{ord}_{\mathcal D_\alpha}(F)$ along each surface.
For example, since $\chi_{10}\sim Z^2$ near $H_1$ ($Z=0$), its zero locus is simply $\mathcal Z(\chi_{10})=H_1$, whereas its divisor is  $\dv(\chi_{10})=2H_1$ on the Siegel-space cover. Denoting by $\phi(\Omega)$ a local modulus parameter (i.e.\ a local coordinate on moduli space) transverse to $\mathcal D_\alpha$ that vanishes linearly on the surface (so that $\mathcal D_\alpha$ is locally $\phi=0$; for instance, $\phi=Z$ for $H_1$), if
$W_0\sim \phi^{-p_0}$ and $H\sim \phi^{-p_H}$ near a component $\phi=0$, then
\begin{equation}
 W_{\rm np}\sim \phi^{-(p_0+p_H)}
 \label{eq:pole-order-physical}
\end{equation}
unless leading terms cancel. Thus, the full superpotential, including possible
numerator zeros, determines whether a proposed vacuum is regular. We reserve
$j(\gamma,\Omega)$ for the automorphy factor, $\mathfrak j(\tau)$ for the elliptic
Klein invariant and $\mathfrak j_1,\mathfrak j_2,\mathfrak j_3$ for the absolute
Igusa invariants defined in Appendix~\ref{app:forms}. The next section determines
the allowed residual functions $H$ for each of the three threshold choices.

%=====================================================================
\section{Complete parametrization of \texorpdfstring{$H(\Omega)$}{H(Omega)}}
\label{sec:param}
%=====================================================================

With the threshold divisor specified in \Cref{sec:setup}, the remaining
problem is to parametrize every weight-zero $H(\Omega)$ without adding a new pole
locus. For Cases (i) and (ii), the holomorphic numerators lie in the even genus-two ring
\begin{equation}
  \Aev = \mathds C[E_4,E_6,\chi_{10},\chi_{12}]\,,
\end{equation}
while $\chi_{35}$ is the odd-weight generator~\cite{Igusa:1962,Igusa:1967}.
When the non-trivial sign $\nu_\chi(\gamma)=\pm1$ is included, holomorphic forms are built using the weight-$5$ form $\chi_5$ (with $\chi_5^2=\chi_{10}$) and the weight-$30$ form $\chi_{30}=\chi_{35}/\chi_5$, whose square $\chi_{30}^2=P_{60}(E_4,E_6,\chi_{10},\chi_{12})$ again lies in $\Aev$ (see Appendix~\ref{app:characters}). Since
$\chi_{10}$ and $\chi_{12}$ have even weight, a weight-zero numerator cannot contain
a single additional factor of $\chi_{35}$. The problem therefore reduces to
modular-weight matching at fixed threshold divisor.

For an even threshold form $\chi$ of weight $k_\chi$, an allowed term has the form
$f/\chi^r$ with $f$ holomorphic of weight $k_\chi r$. A contribution $\chi_{35}g$
would require an even-ring form of odd weight $k_\chi r-35$ and is therefore absent, whereas for $\nu_\chi\neq 1$ the even-weight numerator is proportional to $\chi_{30}$.
The resulting non-negative integer weight equation has finitely many primitive
solutions. The answer may be written either as a pole-order expansion,
\begin{equation}
 H = \sum_{r=0}^{r_{\max}}\frac{P_{k_\chi r}}{\chi^r}\,,
 \label{eq:pole-order-expansion}
\end{equation}
or as a polynomial in the finite sets of basic modular-invariant ratios given below.
The two descriptions are equivalent order by order.

%%%%%%%%%%%%%%%%%%%%%%%%%%%%%%%%%%%%%%%%%%%
\subsection{Case (i): the \texorpdfstring{$\chi_{10}$}{chi10} threshold correction}
\label{sec:chi10}

The allowed monomials based on threshold corrections with $\chi_{10}$ must have the structure
\begin{equation}
 \frac{E_4^aE_6^b\chi_{12}^e}{\chi_{10}^r}\,,
 \qquad 4a+6b+12e=10r\,.
 \label{eq:semigroup-10}
\end{equation}
Every solution is a product of the following eight basic modular-invariant ratios
\begin{equation}
\begin{aligned}
 \mathsf I_1&=\frac{E_4E_6}{\chi_{10}}\,,&\quad
 \mathsf I_2&=\frac{E_4^5}{\chi_{10}^2}\,,&\quad
 \mathsf I_3&=\frac{E_4^2\chi_{12}}{\chi_{10}^2}\,,&\quad
 \mathsf I_4&=\frac{E_6^5}{\chi_{10}^3}\,,\\
 \mathsf I_5&=\frac{E_6^3\chi_{12}}{\chi_{10}^3}\,,&
 \mathsf I_6&=\frac{E_6\chi_{12}^2}{\chi_{10}^3}\,,&
 \mathsf I_7&=\frac{E_4\chi_{12}^3}{\chi_{10}^4}\,,&
 \mathsf I_8&=\frac{\chi_{12}^5}{\chi_{10}^6}\,.
\end{aligned}
\label{eq:Igens}
\end{equation}
Hence,
\begin{equation}
 H^{(10)}(\Omega) ~=~ \mathcal P(\mathsf I_1,\ldots,\mathsf I_8)\,,
 \label{eq:H10}
\end{equation}
where $\mathcal P$ is an arbitrary polynomial,
with the syzygies in Appendix~\ref{app:syzygies} understood. Their ideal is minimally
generated by fifteen binomials. \Cref{eq:H10} is therefore a complete
parametrization of weight-zero functions whose only allowed poles are on
$\chi_{10}=0$.

If $H$ picks up the extra sign $\nu_\chi$, define
\begin{equation}
 \Ich ~=~ \frac{\chi_{30}}{\chi_{10}^3}\,,
 \qquad \chi_{35}~=~\chi_5\chi_{30}\,.
\end{equation}
Since $\chi_{30}^2=P_{60}$, one also has
\begin{equation}
 \Ich^2 ~=~ \frac{P_{60}}{\chi_{10}^6} ~=:~ \widetilde P_{60}(\mathsf I_1\,,\ldots,\mathsf I_8)\,.
 \label{eq:Ichi-closure}
\end{equation}
Then, every allowed function takes the form
\begin{equation}
 H_\nu^{(10)} ~=~ \Ich\,\mathcal P(\mathsf I_1,\ldots,\mathsf I_8)\,.
 \label{eq:H10sign}
\end{equation}

%%%%%%%%%%%%%%%%%%%%%%%%%%%%%%%%%%%%%%%%%%%
\subsection{Case (ii): the \texorpdfstring{$\chi_{12}$}{chi12} threshold correction}
\label{sec:chi12}

Let us now restrict the poles to $\chi_{12}=0$. In this case, the allowed monomials obey
\begin{equation}
 \frac{E_4^aE_6^b\chi_{10}^c}{\chi_{12}^r}\,,
 \qquad 4a+6b+10c~=~12r\,,
 \label{eq:semigroup-12}
\end{equation}
and every solution is generated by the six basic modular-invariant ratios
\begin{equation}
\begin{aligned}
 \mathsf K_1&=\frac{E_4^3}{\chi_{12}}\,,&\qquad
 \mathsf K_2&=\frac{E_6^2}{\chi_{12}}\,,&\qquad
 \mathsf K_3&=\frac{E_4^2E_6\chi_{10}}{\chi_{12}^2}\,,\\
 \mathsf K_4&=\frac{E_4\chi_{10}^2}{\chi_{12}^2}\,,&
 \mathsf K_5&=\frac{E_6\chi_{10}^3}{\chi_{12}^3}\,,&
 \mathsf K_6&=\frac{\chi_{10}^6}{\chi_{12}^5}\,.
\end{aligned}
\label{eq:Kgens}
\end{equation}
Thus, in this case
\begin{equation}
 H^{(12)}(\Omega) = \mathcal Q(\mathsf K_1,\ldots,\mathsf K_6)\,,
 \label{eq:H12}
\end{equation}
where $\mathcal Q$ is an arbitrary polynomial,
with the six minimal syzygies in
Appendix~\ref{app:syzygies} understood.

On $Z=0$, we observe that
\begin{equation}
 \mathsf K_1 = \mathfrak j(T)\,\mathfrak j(U)\,,\qquad
 \mathsf K_2 = [\mathfrak j(T)-1728][\mathfrak j(U)-1728]\,,
 \label{eq:K12-restrictions}
\end{equation}
which will be useful for the examples below.

If $H$ instead transforms with $\nu_\chi$, three extra basic factors are necessary and
sufficient:
\begin{equation}
 L_1 = \frac{E_6\chi_{30}}{\chi_{12}^3}\,,\qquad
 L_2 = \frac{E_4^2\chi_{10}\chi_{30}}{\chi_{12}^4}\,,\qquad
 L_3 = \frac{\chi_{10}^3\chi_{30}}{\chi_{12}^5}\,,
 \label{eq:Lgens}
\end{equation}
so that
\begin{equation}
 H_\nu^{(12)} = \sum_{a=1}^3L_a\,\mathcal Q_a(\mathsf K_1,\ldots,\mathsf K_6)\,,
\end{equation}
where the $\mathcal Q_a$ are arbitrary polynomials.
%%%%%%%%%%%%%%%%%%%%%%%%%%%%%%%%%%%%%%%%%%%
\subsection{Case (iii): threshold corrections involving \texorpdfstring{$\chi_{35}$}{chi35}}
\label{sec:chi35}

In this case, the superpotential we need has the form
\begin{equation}
 W_{\rm np}^{(35)} = \mathcal A\,\e^{-aS}\,\frac{H^{(35)}}{\chi_{35}^p}\,,
 \qquad p \in \mathds N\,,
\end{equation}
where the residual function $H$ has weight zero. Therefore, any additional inverse power of
$\chi_{35}$ appearing in $H$ must occur in pairs. Using
\begin{equation}
 \chi_{35}^2 = \chi_{10}\chi_{30}^2\,,
\end{equation}
the complete result can be written as
\begin{equation}
\label{eq:H35}
 H^{(35)} = \frac{\mathcal P(\mathsf I_1,\ldots,\mathsf I_8)}{\Ich^{2j}}\,,\qquad
 H_\nu^{(35)} = \frac{\mathcal P(\mathsf I_1,\ldots,\mathsf I_8)}{\Ich^{2j+1}}\,,
\end{equation}
where we have used the variables of Case (i). A simple example with the extra sign is
\begin{equation}
 \Gamma_\chi ~=~ \frac{E_6^5}{\chi_{30}} ~=~ \frac{\mathsf I_4}{\Ich}\,.
\end{equation}

%%%%%%%%%%%%%%%%%%%%%%%%%%%%%%%%%%%%%%%%%%%
\subsection{Three absolute Igusa invariants and the complete parametrization}
\label{sec:threeinvariants}

The moduli space has three complex dimensions, so three absolute Igusa invariants
generate its function field rationally. This is weaker than generating all functions
in \Cref{eq:H10,eq:H12} by finite polynomials while keeping the allowed pole divisor fixed: expressing all eight $\mathsf I_{\kappa}$ or six $\mathsf K_{\varsigma}$
rationally in terms of just three invariants typically places $E_4$, $E_6$, or $\chi_{12}$ into the denominators,
introducing apparent poles where the original function is completely regular (see Appendix~\ref{app:adapted}).

At $\OI$ all eight $\mathsf I_{\kappa}$ vanish and none of the fifteen relations contains a
term linear in a single $\mathsf I_{\kappa}$. Hence, eight independent vanishing generators
remain before products of small quantities are taken into account:
\begin{equation}
 \dim_{\mathds C}\frac{\mathfrak m_{\rm I}}{\mathfrak m_{\rm I}^2}=8\,,
 \qquad \mathfrak m_{\rm I}=(\mathsf I_1,\ldots,\mathsf I_8)\,.
 \label{eq:embedding8}
\end{equation}
Likewise, at $\OVI$,
\begin{equation}
 \dim_{\mathds C}\frac{\mathfrak m_{\rm VI}}{\mathfrak m_{\rm VI}^2}=6\,,
 \qquad \mathfrak m_{\rm VI}=(\mathsf K_1,\ldots,\mathsf K_6)\,.
\end{equation}
In contrast, any polynomial ring in just three invariants has at most three independent linear variations around any point.\footnote{Since the three generators of a three-dimensional domain $R$ must be algebraically independent, $R\simeq\mathds C[x_1,x_2,x_3]$ is a free polynomial ring, which has only three independent linear coordinates at every point.}
No choice of three functions can therefore give a polynomially complete parametrization of Cases (i) or (ii). This is the basis-independent
distinction between rational generation and the finite pole-restricted
parametrizations above.

For the standard absolute Igusa functions used in Appendix~\ref{app:forms}, the
comparison is explicit. Combining \Cref{eq:absolute-Igusa} with the Case-(i)
definitions gives
\begin{equation}
\begin{aligned}
 \mathfrak j_1&=2\,3^5\,\mathsf I_8\,,\\
 \mathfrak j_2&=-2^{-3}3^3\,\mathsf I_7\,,\\
 \mathfrak j_3&=2^{-5}3\,\mathsf I_6+2^{-3}3^2\,\mathsf I_7\,.
\end{aligned}
\label{eq:Igusa-in-I}
\end{equation}
The transformation is triangular and invertible, so
\begin{equation}
 \mathds C[\mathfrak j_1,\mathfrak j_2,\mathfrak j_3]
 =\mathds C[\mathsf I_6,\mathsf I_7,\mathsf I_8]
 \subsetneq\mathds C[\mathsf I_1,\ldots,\mathsf I_8]/\Isyz\,.
 \label{eq:Igusa-subring}
\end{equation}
Thus, the standard Igusa triple generates only the subring built from
$\mathsf I_{6,7,8}$; in particular, $H_{(10)}=1+\mathsf I_1/358\,400$ is not polynomial
in that triple. \Cref{eq:embedding8} is the basis-independent statement.

%%%%%%%%%%%%%%%%%%%%%%%%%%%%%%%%%%%%%%%%%%%
\subsection{Comparison with previous \texorpdfstring{$\Spg$}{Sp(4,Z)} stabilization studies}
\label{sec:comparison}

Two recent approaches are particularly relevant. Ref.~\cite{Funakoshi:2024yxg}
uses holomorphic Siegel modular forms in stabilizer-field superpotentials, whereas
Ref.~\cite{Leedom:2026nby} studies the heterotic $STUZ$ system with threshold
corrections, gaugino condensation and a polynomial ansatz in three absolute Igusa
invariants. Our construction is complementary to both. After the threshold
prefactor fixes the required modular weight and pole divisor, we parametrize the
remaining weight-zero function $H$ directly in the full three-complex-dimensional
$\Omega$ sector and do not introduce a stabilizer field.

The difference from a three-Igusa-invariant ansatz is algebraic rather than a choice
of coordinates. Three absolute invariants generate the genus-two modular function
field rationally, but \Cref{eq:embedding8} shows that they cannot generate
polynomially all functions allowed at fixed Case-(i) or Case-(ii) pole divisor. The
eight- and six-invariant descriptions retain precisely this pole information. The
completeness claim made here is therefore restricted to the specified threshold pole
classes, not to arbitrary hidden-sector corrections.

For comparison, a genus-one modular function holomorphic on the upper half-plane and
allowed to have poles only at the cusp is a polynomial in the Klein invariant
$\mathfrak j(\tau)$. Genus two differs at the step from rational generation of the
function field to polynomial generation after a pole divisor has been fixed.

%=====================================================================
\section{Vacuum conditions and elliptic fixed points}
\label{sec:criteria}
%=====================================================================

We now turn to the stabilization of the three complex entries of $\Omega$. The
supersymmetric Minkowski case is especially simple. When both $H$ and its first
derivatives vanish, the masses of the geometric moduli are determined by the
holomorphic Hessian of $H$. We first state this result, then discuss the flat
directions produced by zeros on divisors, and finally impose the additional
constraints at isolated elliptic fixed points.

\subsection{Supersymmetric Minkowski vacua}
\label{sec:vacuum-criterion}

Let us consider the superpotential and K\"ahler potential given by
\begin{equation}
 W=B_S(S)\,\mathcal C(\Omega)\,H(\Omega)\,,
 \qquad
 K=-\gamma_K\log\det Y+K_S(S,\bar S)\,,
 \qquad Y=\im\Omega\,,
 \label{eq:W-local}
\end{equation}
where $\mathcal C(\Omega)$ contains the fixed threshold denominator and is assumed to
be finite and non-zero at the point of interest. If the dilaton is held at
$S=S_\star$, define
\begin{equation}
 \Lambda_S:= K^{S\bar S}
 \left|\partial_S\log B_S+K_S\right|^2_{S_\star}.
 \label{eq:LambdaS}
\end{equation}
This quantity coincides with the function $A(S,\bar S)$ of Refs.~\cite{Leedom:2022zdm,Leedom:2026nby}.
The reduced potential for the geometric moduli is
\begin{equation}
 V_\Omega=\e^K\left[ K^{i\bar\jmath}D_iW D_{\bar{\jmath}} \overline{W} +(\Lambda_S-3)|W|^2\right]\,,
 \qquad i,j\in\{T,U,Z\}\,.
 \label{eq:Vred_short}
\end{equation}

At a regular point of the prefactor, the conditions
\begin{equation}
\label{eq:crit_short}
 H(\Omega_\star)=0
 \qquad\text{and}\qquad
 \dd H|_{\Omega_\star}=0
\end{equation}
are equivalent to
\begin{equation}
 W_\star :=
 \left.W\right|_{\Omega_\star}=0 \qquad\text{and}\qquad \left.D_IW\right|_{\Omega_\star}=0\,,
 \qquad\text{where}\quad I\in\{S,T,U,Z\}\,.
\end{equation}
Thus, the point corresponds to a supersymmetric Minkowski vacuum, independently of the dilaton-dependent value of $\Lambda_S$.
Denote the holomorphic Hessian of $H$ by
\begin{equation}
 \Hess_{ij}(H;\Omega_\star):=
 \left.\partial_i\partial_jH\right|_{\Omega_\star}\,.
 \label{eq:Hessian-H}
\end{equation}
At the critical zero (denoting evaluation at $(S_\star,\Omega_\star)$ henceforth by the subscript $\star$), we have
\begin{equation}
 \left.\partial_i\partial_{\bar\jmath}V\right|_\star
 =\e^{K_\star}\, |B_S\mathcal C|_\star^2\, \Hess_{ik}\,K^{k\bar\ell}_\star\,
  \overline{\Hess}_{\bar\jmath\bar\ell}\,,
 \qquad
 \left.\partial_i\partial_jV\right|_\star=0\,.
 \label{eq:VHshort}
\end{equation}
A rank-three $\Hess$ therefore gives masses to all three complex geometric moduli at
quadratic order. If $\Hess$ has smaller rank, at least one complex direction is
massless at that order. Higher-order terms may still isolate the critical point, so
a rank bound is not a statement about the full potential away from the vacuum.

%%%%%%%%%%%%%%%%%%%%%%%%%%%%%%%%%%%%%%
\subsection{Flat directions from zeros on divisors}
\label{sec:flat-theorem}

A sufficiently high-order zero of the superpotential along a complex surface gives a
simple class of supersymmetric Minkowski vacua. Suppose that a modular form vanishes on a smooth surface $\mathcal D$. Near a regular point of $\mathcal D$, let $\phi(\Omega)$ be the holomorphic local modulus parameter transverse to $\mathcal D$ (introduced in \Cref{sec:setup}) that vanishes linearly on the surface. For the separating surface $H_1$, for example, one may take $\phi=Z$ because $H_1$ is locally $Z=0$.

If the superpotential behaves as
\begin{equation}
\label{eq:divisor-zero-local}
  W = \phi^p F(\Omega)\,,
  \qquad F|_{\mathcal D}\neq0\,,
\end{equation}
then $p\ge2$ implies
\begin{equation}
\label{eq:divisor-Minkowski}
   W=\dd W=0\qquad\text{on }\mathcal D \,.
\end{equation}
Every point of $\mathcal D$ is therefore a supersymmetric Minkowski solution of the
geometric sector. Since $\Siegel$ has three complex dimensions, the two tangent
directions remain exactly flat. For $p=2$ the normal direction can already be
massive at quadratic order; for $p\ge3$ the Hessian also vanishes in the normal
direction.

For a modular form with a multiple zero, only the total local vanishing order matters.
If a modular form $\Phi$ vanishes to order $m$ on $\mathcal D$ and the superpotential contains
$\Phi^r$, then $p=mr$ in \Cref{eq:divisor-zero-local}.
In particular,
\begin{equation}
 \chi_{10}\sim Z^2\qquad\text{near }H_1\,,
\end{equation}
so $\chi_{10}^6\sim Z^{12}$ and corresponds to $p=12$. The relevant local modulus parameter
is therefore $Z$, not $\chi_{10}$ itself.\footnote{Locally near a smooth irreducible component $\mathcal D=\{\phi=0\}$, one has $\Phi=u\,\phi^m$ with $u|_{\mathcal D}\neq0$ and $m=\operatorname{ord}_{\mathcal D}(\Phi)$, so $\Phi^r\sim \phi^{mr}$.}

The same counting applies at genus $g$, where
\begin{equation}
 N_g=\frac{g(g+1)}2
\end{equation}
complex moduli imply $N_g-1$ flat tangent directions on a smooth vacuum divisor.
Genus one is special because the modular space is one-dimensional, so the zero set of
a non-zero modular function is generically discrete.

%%%%%%%%%%%%%%%%%%%%%%%%%%%%%%%%%%%%%%%%%%%
\subsection{Elliptic fixed points and the Hessian}
\label{sec:elliptic-conditions}

We now use the six isolated representatives in \Cref{eq:orbits_short} and their stabilizers in \Cref{tab:fixedall} to
constrain the values and derivatives of the modular forms.
Let $\Omega_\star$ be any $\dim_{\mathds C}=0$ fixed locus, take $\gamma_0$ in its stabilizer
and define $j_0=j(\gamma_0,\Omega_\star)$. A modular form of weight $k$ obeys
\begin{equation}
\label{eq:fixed-value-rule}
  F_k(\Omega_\star) = j_0^k\, F_k(\Omega_\star)\,.
\end{equation}
If $d_\star$ is the order of $j_0$, a form with no additional sign must vanish when
$d_\star$ does not divide $k$. This gives the vanishing pattern in
\Cref{tab:fixed0}. For example, $d_\star=5$ at $\OI$, while $d_\star=12$ at $\OVI$.

For a regular weight-zero invariant function, differentiating
$H(\gamma\Omega)=H(\Omega)$ yields
\begin{equation}
 \partial_\Omega H(\gamma\Omega)
 =(C\Omega+D)\,\partial_\Omega H(\Omega)\,(C\Omega+D)^\top \,.
 \label{eq:grad_short}
\end{equation}
At each isolated elliptic orbit the tangent representation has no invariant vector.
It follows that~\cite{Leedom:2026nby}
\begin{equation}
 \dd H|_{\Omega_\star}=0 \,.
 \label{eq:auto-criticality}
\end{equation}
Thus every regular weight-zero $H$ is critical at these points. The value
$H(\Omega_\star)$ is not fixed by symmetry and must vanish separately for the
supersymmetric Minkowski vacuum of \Cref{eq:crit_short}.

The same argument shows that any smooth modular-invariant scalar potential is
stationary at an isolated elliptic orbit; stationarity alone does not imply stability.
If $W_\star\neq0$, one may instead use the invariant function
$G=K+\log|W|^2$ to obtain $D_iW|_\star=0$. A non-zero $W$ is compatible with the
stabilizer only if its modular transformation law is trivial at the fixed point:
\begin{equation}
 W(\gamma\Omega)=\rho_W(\gamma)j(\gamma,\Omega)^{-\gamma_K}W(\Omega)\,,
 \qquad
 \rho_W(\gamma_0)j_0^{-\gamma_K}=1
 \quad\text{for all }\gamma_0\in G_\star \,.
 \label{eq:W-fixed-phase}
\end{equation}
Here $\rho_W$ denotes a possible one-dimensional phase. If this phase is absent, the
condition reduces to $d_\star\mid\gamma_K$. The condition permits a non-zero
$W_\star$; it does not require one.

Linearizing the modular action leads to
\begin{equation}
 \delta\Omega\longmapsto
 R(\gamma_0)\delta\Omega
 =(C\Omega_\star+D)^{-\top}\delta\Omega\,(C\Omega_\star+D)^{-1}\,.
 \label{eq:tangent-action}
\end{equation}
Since the first derivative vanishes, the Hessian satisfies
\begin{equation}
 R(\gamma_0)^\top\,\Hess\,R(\gamma_0)=\Hess
 \qquad\text{for all }\gamma_0\in G_\star \,.
 \label{eq:Hequiv_short}
\end{equation}
It is therefore enough to determine the symmetric $3\times3$ matrices satisfying
\Cref{eq:Hequiv_short}. We denote their vector space by
\begin{equation}
\label{eq:Qstar}
 \mathcal Q_\star:=  \left[\Sym^2\!\left(T_{\Omega_\star}^*\Siegel\right)\right]^{G_\star}\,,
 \qquad
 n_\star^{(2)}:=\dim_{\mathds C}\mathcal Q_\star\,,
\end{equation}
and define
\begin{equation}
\label{eq:rsym}
  r_\star^{\rm sym} :=  \max_{\Hess\in\mathcal Q_\star}\rank\Hess \,.
\end{equation}
Thus, for any regular weight-zero $H$,
\begin{equation}
\label{eq:rank-bound}
   \rank\Hess(H;\Omega_\star)\le r_\star^{\rm sym}\,.
\end{equation}
\Cref{eq:Qstar} is simply the space of stabilizer-invariant symmetric
bilinear forms on the three complex tangent directions $(\delta T,\delta U,\delta Z)$.
The explicit calculation is given in Appendix~\ref{app:fixedloci}.

\begin{table}[t]
\centering
\small
\begin{tabular}{lcccccccc}
\toprule
orbit & $d_\star$ & $E_4$ & $E_6$ & $\chi_{10}$ & $\chi_{12}$ & $\chi_{35}$
& $n_\star^{(2)}$ & $r_\star^{\rm sym}$\\
\midrule
$\OI$ ($\Z5$)                    &  5 & 0       & 0       & $\neq0$ & 0       & $\neq0$ & 1 & 2\\
$\OII$ ($S_4$)                   &  2 & $\neq0$ & $\neq0$ & $\neq0$ & $\neq0$ & 0       & 1 & 3\\
$\OIII$ ($({\Z4}\x{\Z2})\rtimes{\Z2}$) &  4 & $\neq0$ & 0       & 0       & $\neq0$ & 0       & 1 & 2\\
$\OIV$ ($S_3\x\Z6$)            &  6 & 0       & $\neq0$ & 0       & $\neq0$ & 0       & 0 & 0\\
$\OV$ ($D_{12}$)                 &  2 & $\neq0$ & $\neq0$ & $\neq0$ & $\neq0$ & 0       & 2 & 3\\
$\OVI$ ($\Z{12}$)               & 12 & 0       & 0       & 0       & $\neq0$ & 0       & 1 & 1\\
\bottomrule
\end{tabular}
\caption{Modular-form values and Hessian constraints at the six isolated elliptic
orbits. The last column gives the largest Hessian rank allowed by the stabilizer.
Only the $S_4$ and $D_{12}$ orbits allow rank three for a regular weight-zero $H$.}
\label{tab:fixed0}
\end{table}

For $W=0$, the last column of \Cref{tab:fixed0} directly tests whether all three
complex moduli can be massive at quadratic order. Full rank, i.e.\ rank 3,
is only possible at $\OII$ and $\OV$. The same residual symmetry fixes the multiplicities. At $\OII$,
\begin{equation}
\label{eq:tangent-S4}
   T_{\OII}\Siegel\simeq\mathbf3\qquad(S_4)\,,
\end{equation}
so the canonically normalized mass operator is proportional to the identity on the
complex triplet and all six real geometric scalars are degenerate. At $\OV$,
\begin{equation}
 T_{\OV}\Siegel\simeq
 (\mathbf1'\otimes\mathbf-)\oplus(\mathbf2\otimes\mathbf+)
 \qquad(D_{12}\simeq S_3\times\Z2)\,,
 \label{eq:tangent-D12}
\end{equation}
which gives one complex singlet and one complex doublet, hence a $2+4$ real-mass
pattern. Their mass ratio depends on $H$. The representation-theory decompositions
that determine $n_\star^{(2)}$ and $r_\star^{\rm sym}$ are collected in
Appendix~\ref{app:fixedloci}.

The same rank bounds can be checked directly from the basic ratios. At $\OI$, all
$\mathsf I_{\kappa}$ vanish and only $\mathsf I_1$ contributes at quadratic order, giving
rank at most two. At $\OVI$, all $\mathsf K_{\varsigma}$ vanish and the quadratic term has rank
one. At $\OIII$, the non-zero $\mathsf K_1$ has a rank-two Hessian, while
$\mathsf K_2=E_6^2/\chi_{12}$ starts at fourth order. At $\OIV$, the stabilizer
forbids every invariant quadratic form. These checks agree with
\Cref{tab:fixed0}.

%=====================================================================
\section{Explicit vacua and stabilization examples}
\label{sec:vacua}
%=====================================================================

In this section, we illustrate some sample vacua that can easily be found in our scenario.
In particular, we show that the structure of \Spg leads to continuous and isolated supersymmetric Minkowski vacua, both at and away from fixed points, and that there are stationary points with $W_\star\neq0$ at fixed points, yielding AdS or dS properties.
The detailed study of the phenomenology associated with these solutions and an exhaustive search for AdS/dS vacua is left for future work.

\subsection{Flat-direction benchmarks}

The following three examples illustrate how
zeros of Siegel modular forms can give rise
to continuous families of supersymmetric
Minkowski vacua. For each case, we display
the full superpotential and its regular vacuum
locus, using the notation $\mathcal Z(F)$
introduced in \Cref{eq:zero-locus-def}:
\begin{equation}
\begin{array}{c|c|c}
\text{threshold correction}&W&\text{regular SUSY Minkowski locus}\\ \hline
\chi_{10}
&\mathcal A\e^{-aS}\,\dfrac{\mathsf I_8}{\chi_{10}}
 =\mathcal A\e^{-aS}\,\dfrac{\chi_{12}^5}{\chi_{10}^{7}}
&\mathcal Z(\chi_{12})\setminus\mathcal Z(\chi_{10})\\[8pt]
\chi_{12}
&\mathcal A\e^{-aS}\,\dfrac{\mathsf K_6}{\chi_{12}}
 =\mathcal A\e^{-aS}\,\dfrac{\chi_{10}^6}{\chi_{12}^{6}}
&\mathcal Z(\chi_{10})\setminus\mathcal Z(\chi_{12})\\[8pt]
\chi_{35}
&\mathcal A\e^{-aS}\,\dfrac{\Gamma_\chi}{\chi_{35}}
 =\mathcal A\e^{-aS}\,\dfrac{E_6^5}{\chi_{30}\chi_{35}}
&\mathcal Z(E_6)\setminus\mathcal Z(\chi_{35})
\end{array}
\label{eq:flat-benchmarks}
\end{equation}
On each displayed open surface the numerator has a zero of order at least two in the
normal direction, so $W=\dd W=0$ and the locus is a supersymmetric Minkowski vacuum
manifold. In the second row $\chi_{10}\sim Z^2$ on $H_1$, so the relevant local
order is $p=2\times6=12$ in \Cref{eq:divisor-zero-local}; the criterion of \Cref{sec:flat-theorem} does not
require $\dd\chi_{10}\neq0$. Intersections with a denominator divisor are excluded, since
in such a case the expression is singular or of $0/0$ type and requires a separate local analysis.

The intersection of two regular surfaces also gives a vacuum with one flat complex
direction. In Case (ii), define
\begin{equation}
\label{eq:line-H}
  \widehat P_{60}:=\frac{P_{60}}{\chi_{12}^5} =\frac{\chi_{30}^2}{\chi_{12}^5}\,,
  \qquad
  H_{\rm line}:=\lambda\,\mathsf K_3+\mu\,\widehat P_{60}\,,
 \qquad \lambda\mu\ne0\,,
\end{equation}
where $P_{60}$ is the polynomial in the even generators defined in
Appendix~\ref{app:characters}. Thus, $\widehat P_{60}$ is a polynomial in the allowed
$\mathsf K_{\varsigma}$, and $H_{\rm line}$ introduces no pole away from $\chi_{12}=0$.
With the same regular prefactor as in \Cref{eq:flat-benchmarks},
\begin{equation}
 W_{\rm line} = \mathcal A\e^{-aS}\frac{H_{\rm line}}{\chi_{12}}\,,
 \qquad
 \mathcal L=\left\{\Omega=\begin{pmatrix}\tau&0\\0&\tau \end{pmatrix}:
 \tau\in\mathds H\right\}\subset H_1\cap H_4 \,.
 \label{eq:line-vacuum}
\end{equation}
On $\mathcal L$, both $\chi_{10}$ and $\chi_{30}$ vanish, whereas
$\chi_{12}(\diag(\tau,\tau))=\Delta(\tau)^2\ne0$ at every interior point.
Consequently, $W_{\rm line}=\dd W_{\rm line}=0$ along the entire line.

To count its flat directions, take $x=T-U$ and $z=Z$ transverse to $\mathcal L$.
At a generic $\tau$, the local zeros of $\chi_{30}$ on $H_4$ and of $\chi_{10}$
on $H_1$ have orders one and two, respectively. Hence,
\begin{equation}
 \widehat P_{60}=a(\tau)x^2+\mathcal O\bigl((x,z)^3\bigr)\,,\qquad
 \mathsf K_3=b(\tau)z^2+\mathcal O\bigl((x,z)^3\bigr)\,,
 \qquad a(\tau)b(\tau)\ne0\,,
 \label{eq:line-local}
\end{equation}
where $b(\tau)\ne0$ follows from $\mathsf K_3=E_4^2E_6\chi_{10}/\chi_{12}^2$ whenever
$E_4(\tau)E_6(\tau)\ne0$. The Hessian therefore has rank two at a generic point:
the $x$ and $z$ directions are massive, while the tangent direction
$\delta T=\delta U$, $\delta Z=0$ is exactly flat. At special elliptic values of
$\tau$ the transverse rank can decrease, but the vacuum line persists. This
particular line lies on the gauge-enhancement surface $H_1$, where additional light
states may have to be included in a complete effective theory. A one-dimensional
vacuum family away from $H_1$ follows by choosing a generic point of $H_4$ with
$\chi_{10}\chi_{12}\ne0$, setting $c=\mathsf K_1$ there, and taking
\begin{equation}
 H=\bigl(\mathsf K_1-c\bigr)^2+\mu\,\widehat P_{60}\,,\qquad \mu\ne0\,.
 \label{eq:line-away-H1}
\end{equation}
The locus $H_4\cap\{\mathsf K_1=c\}$ is locally a complex curve when
$\dd\mathsf K_1|_{H_4}\ne0$. At its generic points, the two displayed terms give
non-degenerate quadratic forms in the $H_4$-transverse and $H_4$-normal directions,
respectively, so exactly one complex geometric direction remains flat. As in the
surface examples, a factorized condensate also leaves $S$ flat.

%%%%%%%%%%%%%%%%%%%%%%%%%%%%%%%%%%%%%%%%%%%
\subsection{Isolated fixed-point vacua}

Since any regular weight-zero $H$ satisfies $\dd H|_{\Omega_\star}=0$ at an isolated elliptic orbit, obtaining a supersymmetric Minkowski vacuum at the full-rank points $\OII$ ($S_4$) or $\OV$ ($D_{12}$) requires only the single linear condition $H(\Omega_\star)=0$. Using the exact values in \Cref{tab:gens0}, two simple supersymmetric Minkowski minima are
\begin{equation}
\begin{aligned}
W_{(10)}&=\mathcal A\e^{-aS}\frac{H_{(10)}}{\chi_{10}}\,,
\quad & H_{(10)}&=1+\frac{\mathsf I_1}{358400}\,,\\
W_{(12)}&=\mathcal A\e^{-aS}\frac{H_{(12)}}{\chi_{12}}\,,
& H_{(12)}&=1-\frac{20}{81}\mathsf K_4\,.
\end{aligned}
\label{eq:two_minima}
\end{equation}
At the $S_4$ point $\OII$, the singular values $\sigma(\Hess)$ of the holomorphic Hessian are
\begin{equation}
 \mathsf I_1(\OII)=-358400\,,
 \qquad
 \sigma\!\left(\Hess(H_{(10)};\OII)\right)=(13.9854,6.9927,6.9927)\,,
\end{equation}
so $H=\dd H=0$ and the Hessian has rank three. At the $D_{12}$ point $\OV$,
\begin{equation}
 \mathsf K_4(\OV)=\frac{81}{20}\,,
 \qquad
 \sigma\!\left(\Hess(H_{(12)};\OV)\right)=(24.8840,10.5750,5.0559)\,,
\end{equation}
again giving rank three. In the auxiliary normalization used for the numerical
benchmark, the canonically normalized spectra are
\begin{equation}
 m_{(10)}^2\simeq200.5126(1,1,1,1,1,1)\,,
\end{equation}
\begin{equation}
 m_{(12)}^2=(3.93157,3.93157,12.42570,12.42570,12.42570,12.42570)\,.
 \label{eq:orbit-masses}
\end{equation}
The absolute values are convention dependent; the sixfold and $2+4$ multiplicities
are symmetry predictions.

For reference, Case (iii) also admits a distinct $W_\star\neq0$ mechanism. Taking
\begin{equation}
 W_{(35)} ~=~ \mathcal A\e^{-aS}\frac{1+\mathsf I_1/358400}{\chi_{35}}\,,
\end{equation}
one has $\mathsf I_1(\OI)=0$ and $\chi_{35}(\OI)\neq0$. The total geometric weight
is $-35$ in this benchmark and $d_{\rm I}=5$, so the modular phase condition in
\Cref{eq:W-fixed-phase} permits $W_\star\neq0$ and the fixed-point argument gives $D_iW=0$ at $\OI$.

More generally, at an isolated orbit with $W_\star\neq0$ and $D_iW|_\star=0$, define
\begin{equation}
\label{eq:M-Wnonzero}
   M_{ij} := \nabla_iD_jW|_\star\,,\qquad
   m_{3/2}^2 := \e^{K_\star}|W_\star|^2\,.
\end{equation}
Let $\mu_i\ge0$ denote the singular values of the
canonically normalized complex symmetric matrix
$M$, obtained through its Takagi
factorization\footnote{Every complex symmetric matrix $A$ can be Takagi factorized in terms of a unitary matrix $U$ and a diagonal matrix
$D=\operatorname{diag}(\mu_1,\ldots,\mu_n)$ as $A=U D U^\top$.}~\cite{Dieci:2022takagi}.
The six squared masses of the real geometric moduli are\footnote{In K\"ahler normal coordinates,
$V_{i\bar\jmath}=\e^{K_\star}
[(MM^\dagger)_{i\bar\jmath}
+(\Lambda_S-2)|W_\star|^2\delta_{i\bar\jmath}]$
and
$V_{ij}=\e^{K_\star}(\Lambda_S-1)
\overline W_\star M_{ij}$,
which leads to \Cref{eq:Wnonzero-masses}
after Takagi factorization.}
\begin{equation}
\label{eq:Wnonzero-masses}
  m_{i,\pm}^2 = \e^{K_\star}\left(\mu_i\pm|W_\star|\right)
  \left[\mu_i\pm(\Lambda_S-2)|W_\star|\right]\,.
\end{equation}
At $\Lambda_S=3$ this becomes
\begin{equation}
 m_{i,\pm}^2 = \e^{K_\star}\left(\mu_i\pm|W_\star|\right)^2\,.
 \label{eq:Z5-masses}
\end{equation}
The residual $\Z5$ symmetry gives $\rank M\le2$; hence, at least one $\mu_i=0$. For
such a singular value, the corresponding two real scalars satisfy
\begin{equation}
\label{eq:mass-zero-holomorphic}
   m_{i,+}^2 = m_{i,-}^2 = (\Lambda_S-2)m_{3/2}^2 = m_{3/2}^2+V_\star\,,
   \qquad V_\star = (\Lambda_S-3)m_{3/2}^2\,.
\end{equation}
Thus, a vanishing holomorphic mass does not imply a flat real scalar direction when
$W_\star\neq0$. For $\Lambda_S=3$ all masses in \Cref{eq:Wnonzero-masses} are non-negative.
Away from this value, stability follows directly from the same formula.
%The numerical scan around the benchmark coefficient is therefore only a check that no accidental zero mode is present.
A complete model must still stabilize $S$ and test the mixed Hessian.

%%%%%%%%%%%%%%%%%%%%%%%%%%%%%%%%%%%%%%%%%%%
\subsection{Vacua away from fixed points and near-critical stabilization}
\label{sec:nearcritical}

Residual symmetry simplifies the fixed-point analysis, but an isolated vacuum need
not lie at a fixed point. At a generic point, a supersymmetric Minkowski vacuum is
obtained whenever $H=\dd H=0$ and the Hessian has rank three. A strict-interior
example in Case (ii) is
\begin{equation}
\Omega_{\rm int}=\begin{pmatrix}
-0.03+1.11470054\I&0.06+0.54735027\I\\
0.06+0.54735027\I&0.04+1.18470054\I
\end{pmatrix},
\label{eq:Omega-int}
\end{equation}
with
\begin{equation}
 H_{\rm int}=\sum_{\varsigma\in\{1,2,4\}}
 \left(\frac{\mathsf K_{\varsigma}}{\mathsf K_{\varsigma}(\Omega_{\rm int})}-1\right)^2\,.
\end{equation}
Then $H_{\rm int}=\dd H_{\rm int}=0$ and
\begin{align}
 \sigma\!\left(\Hess(H_{\rm int};\Omega_{\rm int})\right)
 &= (25.9481,21.8570,1.52545)\,,\nonumber\\
 m_{\rm int}^2
 &= (0.217129,0.217129,3.964676,3.964676,159.874276,159.874276)\,.
\label{eq:int-masses}
\end{align}
All Gottschling inequalities are strict and Newton iteration returns to the target
from small displacements. Thus, full stabilization is not tied to a residual
symmetry, although the fixed-point solutions require less tuning.

A second construction determines the vacuum point from the criticality equations
rather than inserting it into the coefficients. In Case (ii) consider
\begin{equation}
 H=\mathsf K_1+c\mathsf K_2+c_0\,.
 \label{eq:H4-linear-family}
\end{equation}
The gradient equation requires the two one-forms to become linearly dependent,
\begin{equation}
 \Sigma=\{\Omega\in\Siegel:\ \dd \mathsf K_1\wedge\dd \mathsf K_2=0\}\,.
 \label{eq:Sigma}
\end{equation}
The appearance of a curve on $H_4$ follows from its residual $\Z2$ symmetry.
The tangent directions transform as $(+,+,-)$, so an invariant function has no first
derivative along the odd normal direction. On the two-complex-dimensional surface
$H_4$, the condition $\dd\mathsf K_1\wedge\dd\mathsf K_2=0$ reduces to one independent
equation and generically defines a complex curve. Numerically, $\Sigma$ contains
such a curve. On the $\Spg$-equivalent boundary $2Z=T$ of $\mathcal F_2$, one representative is
\begin{equation}
\Omega_c=\begin{pmatrix}
-0.108392+1.020092\I&-0.054197+0.510046\I\\
-0.054197+0.510046\I&0.075903+1.224382\I
\end{pmatrix},
\qquad 2Z=T\,,
\label{eq:Omega-c}
\end{equation}
with
\begin{equation}
 c=1.6084188-1.0214827\I\,,
 \qquad
 c_0=-\bigl(\mathsf K_1+c\mathsf K_2\bigr)\big|_{\Omega_c}\,.
\end{equation}
After normalizing $H$ by $|\mathsf K_1(\Omega_c)|$, the Hessian and reduced mass eigenvalues are
\begin{equation}
 \sigma\!\left(\Hess(H;\Omega_c)\right)=(112.7,12.99,6.844)\,,
 \qquad
 m^2=(5.081,16.94,559.5)\,,
 \label{eq:H4-spectrum}
\end{equation}
with each real eigenvalue doubly degenerate. Here, both $\Omega_c$ and the relative
coefficient $c$ are outputs of the criticality equations.

We can also use the flavor fits discussed in \Cref{sec:siegelspace} as
targets for the stabilization potential. In the restricted $Z=0$ fit of Ref.~\cite{Ding:2020zxw},
\begin{equation}
 \tau_\ell=0.01541+1.00011\I\,,
 \qquad \Omega_A=\diag(\tau_\ell,\tau_\ell)\,,
\end{equation}
which lies only $0.02179$ from $\OIII$. With $\mathfrak j_\ell:=\mathfrak j(\tau_\ell)$ and
$\mathsf K_{\varsigma}^\star:=\mathsf K_{\varsigma}(\Omega_A)$, using \Cref{eq:K12-restrictions} leads to
\begin{equation}
 H_A=\frac{\mathsf K_1}{\mathsf K_1^\star}+c_\ell\frac{\mathsf K_2}{\mathsf K_2^\star}-(1+c_\ell)\,,
 \qquad
 c_\ell=-\frac{\mathfrak j_\ell-1728}{\mathfrak j_\ell}
       =0.00341667+0.00010165\I \,,
\end{equation}
and one finds $H_A=\dd H_A=0$ with
\begin{equation}
 \sigma\!\left(\Hess(H_A;\Omega_A)\right)=(57.5934,57.5934,9.5990)\,.
\end{equation}

\begin{figure}[t]
\centering
\includegraphics[width=0.82\textwidth]{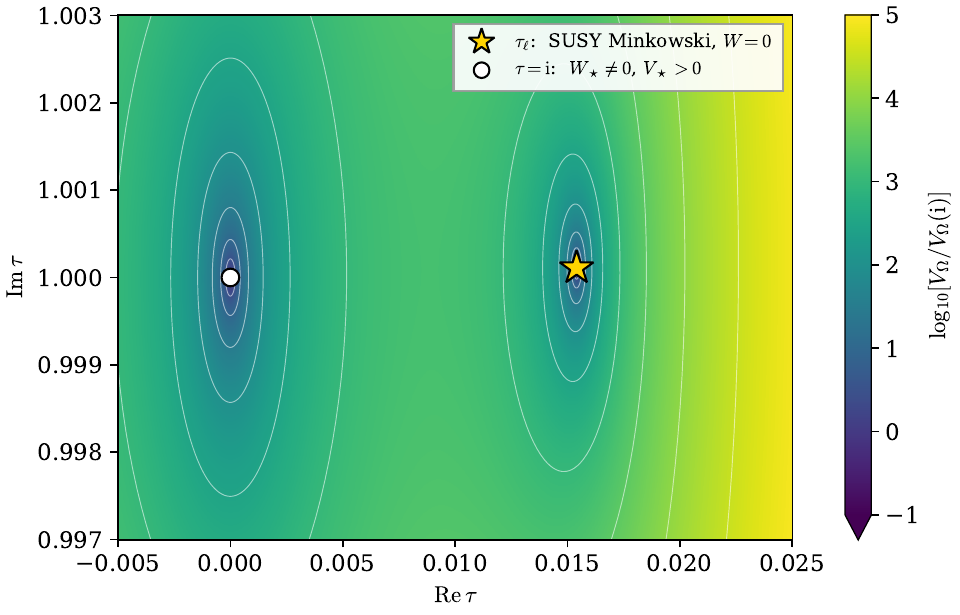}
\caption{\label{fig:offfixed-heatmap}
Reduced Case-(ii) potential on the slice $Z=0$, $T=U=\tau$, using the
auxiliary normalization $\gamma_K=12$ and $\Lambda_S=4$, shown as
$\log_{10}[V_\Omega/V_\Omega(\I)+10^{-12}]$. The star marks the off-fixed
supersymmetric Minkowski vacuum $\tau_\ell=0.01541+1.00011\,\I$ and the circle marks
the $W_\star\neq0$ positive-energy elliptic stationary point $\tau=\I$. The overall
condensate scale has been divided out. The plot is generated directly from the
genus-one restrictions of \Cref{eq:K12-restrictions}.}
\end{figure}

\Cref{fig:offfixed-heatmap} visualizes this example. Using the same auxiliary
Case-(ii) normalization $\gamma_K=12$, we set $\Lambda_S=4$ in the reduced
potential~\eqref{eq:Vred_short} and restrict to the two-real-dimensional slice $Z=0$,
$T=U=\tau$. On this slice $\mathsf K_1=\mathfrak j(\tau)^2$,
$\mathsf K_2=[\mathfrak j(\tau)-1728]^2$ and $\chi_{12}=\Delta(\tau)^2$. The supersymmetric Minkowski minimum is at $\tau_\ell$, not at $\tau=\I$. At $\tau=\I$, however, $H_A(\OIII)=c_\ell(1+c_\ell)\neq0$.
Since $d_{\rm III}=4$ divides $\gamma_K=12$,
this is a distinct stationary point with $W_\star\neq0$. For the illustrative choice $\Lambda_S=4$, its vacuum energy is positive:
$V_\star = m_{3/2}^2 > 0$. Choosing $\Lambda_S<3$ would instead give
a negative vacuum energy, but either case requires
independent tests of dilaton stationarity
and stability in the full theory.

The figure shows only
a slice of the full six-real-dimensional potential. The full-rank Hessian at
$\tau_\ell$ is the independent local test of isolation, while stability of the
elliptic $W_\star\neq0$ point is governed by \Cref{eq:Wnonzero-masses}.

Let us now consider a second flavor point that lies on $H_4$ and close to $Z=1/2$:
\begin{equation}
 \Omega_B=\begin{pmatrix}
0.04017+0.89185\I&0.49053+0.00792\I\\
0.49053+0.00792\I&0.04017+0.89185\I
\end{pmatrix}.
\end{equation}
Writing $q_{\varsigma}=\mathsf K_{\varsigma}/\mathsf K_{\varsigma}(\Omega_B)$, the choice
\begin{equation}
 H_B=c_1q_1+c_2q_2+q_4-(c_1+c_2+1)\,,
\end{equation}
with
\begin{equation}
 c_1=-3.63575318+0.08170352\I\,,\qquad
 c_2=-1.68840231-0.04125274\I\,,
\end{equation}
gives $H_B=\dd H_B=0$ and
\begin{equation}
 \sigma\!\left(\Hess(H_B;\Omega_B)\right)=(309.575,4.7673,0.05328)\,.
\end{equation}
The hierarchy is large, but proximity to $Z=1/2$ alone does not protect the smallest eigenvalue.

A near-critical family of isolated supersymmetric Minkowski vacua can be constructed from the vacuum
curve in \Cref{eq:line-vacuum}. Consider
\begin{equation}
\label{eq:nearcritical-vacuum}
 H_v=\lambda\,\mathsf K_3+\mu\,\widehat P_{60}
 +\kappa\left[\mathsf K_1-\mathfrak j(\tau_v)^2\right]^2\,,
 \qquad
 \Omega_v=\diag(\tau_v,\tau_v)\,,
\end{equation}
where $\lambda,\mu,\kappa$ are fixed non-zero constants,
while $\tau_v$ is a point in moduli space that may be considered as a label of different superpotentials.
Since at $Z=0$ we have
$\widehat P_{60}
\propto[\mathfrak j(T)-\mathfrak j(U)]^2$ and $\mathsf K_1(\Omega_v)=\mathfrak j(\tau_v)^2$,
we find $H_v=\dd H_v=0$ at $\Omega_v$.
By taking the derivatives of $H_v$, one can show that, for generic $\tau_v$, the Hessian has full rank, with its three quadratic terms proportional to
$\mathfrak j'(\tau_v)^2$,
$\mathfrak j'(\tau_v)^2$ and $E_6(\tau_v)^2$,
respectively, along the directions
$\delta T+\delta U$, $\delta T-\delta U$
and $\delta Z$.
Here $E_6(\tau)$ denotes the genus-one Eisenstein
series. Since both $\mathfrak j'(\tau)$ and
$E_6(\tau)$ vanish linearly at $\tau=\I$,
the geometric moduli masses satisfy
(cf.~\Cref{eq:VHshort})
\begin{equation}
\label{eq:light_short}
 \sigma_i=\mathcal O(d^2)\,,\qquad
 m_i^2=\mathcal O(d^4)\,,\qquad
 i=1,2,3\,,\qquad d=|\tau_v-\I|\,.
\end{equation}
Thus, all three geometric moduli become light
as the vacuum approaches $\OIII$, relative to
the fixed overall superpotential scale.

%%%%%%%%%%%%%%%%%%%%%%%%%%%%%%%%%
\subsection{Dilaton stabilization}

The $W=0$ examples stabilize $\Omega$ but leave $S$ flat whenever the $S$ dependence factorizes. If
\begin{equation}
 W=B_S(S)F(\Omega)\,,\qquad
 F(\Omega_\star)=\dd F|_\star=0\,,
\end{equation}
then
\begin{equation}
  W = \partial_IW = \partial_S^2W = \partial_S\partial_iW = 0
\end{equation}
for every $S$. Thus, even an arbitrary factorized racetrack cannot lift the dilaton.

To stabilize $S$, this complete factorization must be broken. One possibility is to
allow different condensates to have different $\Omega$ dependence, as in
\begin{equation}
\label{eq:multi_short}
   W=\frac1{\chi^n}\sum_{\alpha=1}^{N} A_\alpha\e^{-a_\alpha S}H_\alpha(\Omega)\,.
\end{equation}
At an elliptic point, every regular weight-zero $H_\alpha$ is individually critical.
Writing
\begin{equation}
  v_\alpha = A_\alpha\,\e^{-a_\alpha S_\star}\,H_\alpha(\Omega_\star)\,,
\end{equation}
the conditions $W=D_SW=0$ become
\begin{equation}
 \sum_\alpha v_\alpha=0\,,\qquad
 \sum_\alpha a_\alpha v_\alpha=0\,.
\end{equation}
Two terms with different $a_\alpha$ force the trivial solution. Three are the minimal
possibility, with
\begin{equation}
 (v_1,v_2,v_3)=\lambda(a_2-a_3,a_3-a_1,a_1-a_2)\,.
 \label{eq:v3_short}
\end{equation}
Then, the mixed holomorphic $S$--$\Omega$ block vanishes in the separable reduced K\"ahler setup, while
\begin{equation}
 \nabla_iD_jW|_\star =\chi(\Omega_\star)^{-n}\sum_\alpha A_\alpha\e^{-a_\alpha S_\star}
 \partial_i\partial_jH_\alpha|_\star\,.
\end{equation}
The residual-symmetry mass multiplicities therefore survive. Moreover
\begin{equation}
 \partial_S^2W|_\star =\chi(\Omega_\star)^{-n}\sum_\alpha a_\alpha^2v_\alpha
                      =-\lambda\chi(\Omega_\star)^{-n} (a_1-a_2)(a_2-a_3)(a_3-a_1)\,,
\end{equation}
which is non-zero for distinct exponents. In the reduced separable K\"ahler setup,
\begin{equation}
 m_S^2 = \e^{K_\star}\left(K^{S\bar S}_\star\right)^2\,
 \left|\frac{\sum_\alpha a_\alpha^2v_\alpha}{\chi(\Omega_\star)^n}\right|^2\,.
\end{equation}
At $\OII$ the geometric Hessian remains proportional to the unique $S_4$-invariant
quadratic form, so the sixfold degeneracy is preserved. At $\OV$ the $2+4$ pattern
survives, although the splitting can change. This is an effective completion, not a
claim that a generic compactification supplies the required three condensates and
coefficient ratios. Note also that a Green--Schwarz-mixed K\"ahler potential can reintroduce $S$--$\Omega$ mixing.

%=====================================================================
\section{Physical implications and scope}
\label{sec:discussion}
%=====================================================================

The analysis above separates the geometric information fixed by $\Spg$ from the overall mass scale, which remains
compactification dependent. The threshold correction fixes the allowed singular
divisor, while the vacuum geometry determines exact flat directions and constrains
the quadratic spectrum. The supersymmetric Minkowski examples have $W_\star=0$ and
$m_{3/2}=0$; supersymmetry breaking and the absolute mass scale must therefore come
from the remaining hidden sector.

At elliptic stationary points with $W_\star\neq0$, residual symmetry instead
constrains $M_{ij}=\nabla_iD_jW$. A symmetry-forbidden holomorphic quadratic term
need not give a massless real scalar because supergravity contributes
$m_{3/2}^2+V_\star$, as in \Cref{eq:mass-zero-holomorphic}. At $\OIV$, the
stabilizer forces $M_{ij}=0$ in the reduced geometric sector, so all six real
geometric scalars obey this relation; at $\OIII$ and $\OVI$ the symmetry guarantees
at least one and two such complex modes, respectively. Dilaton mixing can modify
these relations. This branch is not needed for the Minkowski vacua that are the main focus here.

\paragraph{Inflation.}
The positive-dimensional fixed sets of $\Spg$ distinguish
directions tangent and normal to the fixed locus. If the normal directions are
stabilized more strongly, a lighter tangent direction can provide an effectively
single-field trajectory without imposing a truncation by hand. Along
$T=U=\tau$, $Z=\tau/2$,
\begin{equation}
 \det\im \Omega = \frac34(\im\tau)^2\,,\qquad
 K = -2\gamma_K\log\im\tau+\text{const.}\,,
\end{equation}
which corresponds to $\alpha_{\rm eff}=2\gamma_K/3$ in the usual
$\alpha$-attractor normalization. This connects directly with recent
$\Spg$ modular-inflation models~\cite{Jiang:2025qbi}. Light geometric moduli can also arise near $\OIII$, as in the example of
\Cref{eq:light_short}, although there all three masses are suppressed together. Whether such a direction supports slow roll depends on
higher derivatives of the potential and on the full hidden-sector dynamics.

\paragraph{Axions and light moduli.}
A continuous supersymmetric Minkowski divisor contains exactly flat tangent
directions in the reduced geometric sector. Such a direction may provide a starting
point for an axion or Peccei--Quinn sector, but flatness alone is insufficient. The
saxion must be stabilized, the QCD anomaly coupling must be present, and additional
sources of symmetry breaking must remain small~\cite{Buchbinder:2014qca}. On the
separating surface $H_1$, for example,
\begin{equation}
   K_{T\bar T} = \frac{\gamma_K}{4(\im T)^2}\,.
\end{equation}
If $\re T$ has unit period and the QCD angle contains $2\pi q_T\re T$, the geometric
normalization gives
\begin{equation}
   f_T \sim \frac{\sqrt{\gamma_K}M_{\rm Pl}}{2\sqrt2\pi|q_T|\im T}\,,
\end{equation}
up to the normalization of the gauge kinetic function. %For order-one moduli this scale is naturally high.
For values of $\im T$ and $q_T$ of order one,
this gives a decay constant of order
$M_{\rm Pl}$, up to numerical factors.

The near-critical mechanism gives another route to a light field. Close to an
enhanced-symmetry point, the geometric modulus masses can be suppressed, as illustrated by the vacuum family defined by \Cref{eq:nearcritical-vacuum}. After supersymmetry breaking the \CP-even and \CP-odd components need not
remain degenerate. A long-lived pseudoscalar may behave as axion-like dark matter,
while an ultralight scalar could enter dark-energy models. The present construction
accounts only for a suppression relative to the overall superpotential scale. It does not determine
the very small absolute scale required for dark energy or remove the usual fifth-force
and cosmological-stability constraints.

\paragraph{Connection with modular flavor.}
In models with modular symmetries, fixed loci in moduli space can exhibit enhanced residual symmetries, leading to the phenomenon of \emph{local flavor unification}~\cite{Baur:2019kwi,Baur:2019iai}. Following the seminal work by Feruglio~\cite{Feruglio:2017spp}, the related mechanism of \emph{near-critical behavior} has been extensively investigated in both one-modulus and multi-moduli frameworks to achieve realistic phenomenology~\cite{Feruglio:2021dte,Novichkov:2021evw,Baur:2021bly,Baur:2022hma,Feruglio:2023prl,Ding:2024xhz,Chen:2025mfh}. In this case, small departures from the symmetric loci can give rise to realistic fermion mass hierarchies.

A similar role of residual symmetry appears in moduli
stabilization. Here, it acts on the geometric fluctuations
$(\delta T,\delta U,\delta Z)$ rather than directly on
fermion multiplets: the residual group constrains the
superpotential Hessian, while in flavor models it
constrains fermion mass matrices. Although the relevant
representations differ, both are governed by the same
underlying modular symmetry. In particular, the $S_4$
and $D_{12}$ vacua exhibit sixfold and $2+4$ real
modulus-mass degeneracies, respectively. Near $\OIII$,
the symmetry-forbidden quadratic term of a fixed function $H$ is gradually
restored, leading to a Hessian singular value that
scales quadratically with the displacement $d$, in close analogy with fermion mass matrices in a fixed modulus background. For moduli masses,
the Hessian must in addition be evaluated at a critical point of the same function; in the vacuum family of
\Cref{eq:nearcritical-vacuum} this leads to $m_i^2=\mathcal O(d^4)$.

Finally, modular-flavor studies often focus on symmetric subspaces, such as $T=U$~\cite{Ding:2020zxw} or
$Z=0$~\cite{Baur:2020jwc,Baur:2021mtl,
Carducci:2026shs}, without addressing their dynamical stabilization. Our examples in \Cref{sec:nearcritical}
show that modulus values favored by existing flavor
fits can also be realized as isolated supersymmetric
Minkowski minima in the full three-complex-dimensional geometric moduli space, without restricting fluctuations to these subspaces.

\paragraph{Microscopic limitations.}
The $\chi_{10}$ divisor has the clearest perturbative interpretation because it is
associated with additional light states. The microscopic meaning of the
$\chi_{12}$ divisor and of an isolated $\chi_{35}$ pole is more compactification
dependent. Likewise, the numerical masses in \Cref{sec:vacua} use a reduced
K\"ahler normalization. The robust results are the existence and rank of the critical
points and the symmetry-imposed mass multiplicities.

The threshold divisor fixes the singular locus, whereas $H$ can change or cancel the
local pole order without moving that locus. When numerator and denominator vanish
simultaneously, the full superpotential must therefore be analyzed from their local
orders, as in \Cref{eq:flat-benchmarks}.

The $W=0$ vacua have $V_\star=0$, so the dimensionless ratios entering the original
and refined de Sitter conjectures are not defined at the vacuum~\cite{Obied:2018sgi,Ooguri:2018wrx}.
Positive-energy continuations require the full dilaton and supersymmetry-breaking sectors
\cite{Gonzalo:2018guu,Leedom:2022zdm}.
In terms of the local modulus parameter $\phi(\Omega)$ vanishing linearly on the threshold divisor, an uncanceled pole in the threshold factor $\Psi(\Omega)$ still gives the local barrier
\begin{equation}
  V_F\sim |\phi|^{-2(rn+1)} = |\Psi|^{-2n-2/r}\to+\infty
\end{equation}
when $\Psi\sim \phi^r$ and $W\sim\Psi^{-n}$. For $\chi_{10}=0$ this is also the region
where additional perturbative states can invalidate the Wilsonian description. For
the other divisors the same scaling holds algebraically, but its microscopic
interpretation is model dependent.

%=====================================================================
\section{Conclusions}
\label{sec:conclusions}
%=====================================================================

By expanding the $\SL{2,\ZZ}$ moduli stabilization framework to the genus-two Siegel modular symmetry \Spg, the moduli space is enlarged to three complex dimensions, incorporating the complex structure, the K\"ahler modulus, and a Wilson-line modulus in the $2\x2$ period matrix $\Omega$.  This gives rise to a richer set of available
cusp forms $\left( \chi_{10},\, \chi_{12},\,\chi_{35}\right)$ and fixed loci in moduli space. As in the simpler \SL{2,\ZZ} case, gaugino condensation can help to dynamically stabilize the moduli, inducing non-perturbatively a superpotential term proportional to an \Spg-invariant function $H(\Omega)$. We have shown that naive \SL{2,\ZZ}-inspired techniques to
construct the general weight-zero function $H$ based on polynomials of the modular absolute invariants are insufficient to generate all functions $H$ compatible with a given singularity structure of the superpotential.
Since the threshold contributions are determined by the \Spg cusp forms, in \Cref{sec:param} we have derived for each of them the complete finite parametrization of the corresponding $H$ function in terms of the basic modular-invariant ratios together with their syzygies. We emphasize that the superpotential contributions associated with thresholds depending on $\chi_{10}$ and $\chi_{35}$ can be parametrized using the same basic modular-invariant ratios, whereas the $\chi_{12}$ case requires a different set of ratios.

An important physical consequence of this exact parametrization is the emergence of a richer vacuum structure.
Most notably, this framework reveals novel supersymmetric Minkowski vacua along zero loci of modular forms in the moduli space. This includes complex
dimension-2 hypersurfaces of supersymmetric Minkowski vacua, which have no analogue in \SL{2,\ZZ} models and represent a distinct
phenomenological feature of the \Spg symmetry.
We also constructed complex dimension-1 vacuum curves with one flat and two massive complex directions, including examples away from $H_1$. At the six isolated elliptic orbits, modular
invariance makes a regular weight-zero $H$ automatically critical, while the residual stabilizer
restricts the quadratic Hessian. Only the $S_4$ orbit $\OII$ and the $D_{12}$ orbit
$\OV$ permit stabilization of all \Spg moduli at supersymmetric Minkowski minima. Additionally, we constructed
explicit isolated vacua away from zero-dimensional fixed points, including vacua at modulus values favored by existing modular flavor models. Our near-critical analysis further provides an explicit family of isolated supersymmetric Minkowski vacua whose geometric
squared masses scale as $\mathcal O(d^4)$, where $d$ measures the
displacement from the fixed point $\OIII$.
We also studied stationary points with $W_\star\neq0$, whose vacuum energy can be positive or negative. Establishing fully stable AdS or dS vacua requires a separate analysis of the dilaton sector. This task along with an extensive search of such vacua is left for future work.

With the full parametrization and vacuum criteria developed here, we have provided a direct starting point for addressing the top-down problems of deriving the threshold divisor, pole order, coefficients of the $H$ function, and hidden-sector dynamics from specific heterotic compactifications toward studying moduli stabilization and its phenomenology.
The exact same unbroken modular symmetries (such as $S_{4}$ and $D_{12}$)
that restrict the superpotential Hessian are also known to restrict fermion mass matrices in the matter
sector, making our off-fixed numerical constructions a useful first step for top-down models seeking to dynamically
select specific flavor structures. Likewise, the light moduli of \Cref{eq:light_short} could serve
as dark matter candidates once their coupling to hidden-sector dynamics is specified. Finally, since
the constructed supersymmetric Minkowski vacua yield $V_\star=0$ and lie outside the scope of the refined de Sitter conjectures, incorporating explicit supersymmetry breaking or uplifting
to transition these Minkowski points into de Sitter vacua provides a natural target for future work.

\paragraph{Acknowledgments.}
We thank Gerard van der Geer for helpful correspondence.  X.-G.L., H.P.N. and S.R.-S.\ are grateful to the Mainz Institute for Theoretical
Physics (MITP) of the Cluster of Excellence PRISMA+ (Project ID 390831469) for its hospitality and support. This work was partly supported by UNAM-PAPIIT IN117226. X.-G.L.\ acknowledges support from the Universidad Nacional
Aut\'onoma de M\'exico Postdoctoral Program (POSDOC).
We acknowledge the use of Claude and ChatGPT for editorial
assistance and some calculations, with all results independently verified by the authors.

\clearpage

\appendix
%=====================================================================
\section{Useful genus-two modular forms and their zero loci}
\label{app:forms}

Igusa's structure theorems give the even-weight genus-two ring as the polynomial ring
generated by $E_4,E_6,\chi_{10},\chi_{12}$ of weights $4,6,10,12$
\cite{Igusa:1962}, and the full ring with no extra $\nu_\chi$ sign by adjoining the
weight-$35$ cusp form $\chi_{35}$, with $\chi_{35}^2\in\Aev$ \cite{Igusa:1967}.
Standard treatments are Refs.~\cite{Klingen:1990,Freitag:1983,vanderGeer:2008}.
These forms, together with the weight-$5$ form $\chi_5$ in the non-trivial
one-dimensional representation, have explicit theta-constant representations
\cite{Igusa:1964,Mumford:1983}, which we use for all numerics.

\subsection{Theta-constant representations}
\label{app:theta}

For a characteristic $m=(m',m'')$, with $m'=(m_1,m_2)$, $m''=(m_3,m_4)$ and $m_i\in\{0,1\}$,
the genus-2 theta constant is defined by
\begin{equation}
\theta_m(\Omega)=\sum_{n\in\ZZ^2}\exp\Bigl[\pi\I\bigl(n+\tfrac12m'\bigr)^{\!\top}\Omega\bigl(n+\tfrac12m'\bigr)+\pi\I\bigl(n+\tfrac12m'\bigr)^{\!\top}m''\Bigr]\,.
\end{equation}
The characteristic is even if $m'\cdot m''=0\bmod 2$ and odd otherwise; odd theta
constants vanish identically. The ten even characteristics, written as
$m_1m_2m_3m_4$, are $0000$, $0001$, $0010$, $0011$, $0100$, $0110$, $1000$, $1001$,
$1100$ and $1111$.
In terms of these theta constants, define
\begin{equation}
\begin{aligned}
X_1&:=\frac{\theta_{0000}^4+\theta_{0001}^4+\theta_{0010}^4+\theta_{0011}^4}{4}\,,\quad
&X_2&:=(\theta_{0000}\theta_{0001}\theta_{0010}\theta_{0011})^2\,,\\
X_3&:=\frac{(\theta_{0100}^4-\theta_{0110}^4)^2}{16384}\,,
&X_4&:=\frac{(\theta_{0100}\theta_{0110}\theta_{1000}\theta_{1001}\theta_{1100}\theta_{1111})^2}{4096}\,,
\end{aligned}
\end{equation}
so that, in the convention of Ref.~\cite{Jiang:2025qbi},
\begin{subequations}
\begin{align}
E_4&=4 X_1^2 - 3 X_2 + 12288 X_3\,,\\
E_6&=-8 X_1^3 + 9 X_1 X_2 + 73728 X_1 X_3 - 27648 X_4\,,\\
\chi_{10}&=-\tfrac14X_2X_4=-2^{-14}\prod_{m\,\rm even}\theta_m^2\,,\\
\chi_{12}&=\tfrac1{12}\bigl(3X_2^2X_3-2X_1X_2X_4+3072X_4^2\bigr)\,,\\
\chi_5&=\I\,2^{-7}\prod_{m\,\rm even}\theta_m\,,\qquad\qquad \chi_5^2=\chi_{10}\,,
\end{align}
\end{subequations}
with $E_4,E_6$ normalized to $1$ at the cusp. These conventions yield
\begin{equation}
\begin{aligned}
E_4\big|_{Z=0}&=E_4(T)E_4(U)\,,\qquad&
E_6\big|_{Z=0}&=E_6(T)E_6(U)\,,\\
\chi_{10}\big|_{Z=0}&=0\,,\quad&
\chi_{12}\big|_{Z=0}&=\Delta(T)\Delta(U)\,,
\end{aligned}
\label{eq:Z0-restrictions}
\end{equation}
with $E_4,E_6,\Delta(\tau):=\eta(\tau)^{24}$ on the right the usual genus-one forms.
\Cref{eq:Z0-restrictions} leads to \Cref{eq:K12-restrictions} by using
$\mathfrak j=E_4^3/\Delta$ and $\mathfrak j-1728=E_6^2/\Delta$.

\subsection{Fourier expansions}
\label{app:fourier}

We record the expansions in $q_1=\e^{2\pi\I T}$, $q_2=\e^{2\pi\I U}$ and
$q_3=\e^{2\pi\I Z}$, generated in exact integer arithmetic from the theta series and
used throughout as a normalization check. Through total degree three in
$(q_1,q_2)$,
\begin{subequations}
\begin{align}
E_4&=1+240(q_1+q_2)+2160(q_1^2+q_2^2)+6720(q_1^3+q_2^3)
\nonumber\\
&\quad+240\bigl(q_3^{-2}+56q_3^{-1}+126+56q_3+q_3^{2}\bigr)q_1q_2
\nonumber\\
&\quad+240\bigl(126q_3^{-2}+576q_3^{-1}+756+576q_3+126q_3^{2}\bigr)(q_1^2q_2+q_1q_2^2)+\ldots,
\\
E_6&=1-504(q_1+q_2)-16632(q_1^2+q_2^2)-122976(q_1^3+q_2^3)
\nonumber\\
&\quad-504\bigl(q_3^{-2}-88q_3^{-1}-330-88q_3+q_3^{2}\bigr)q_1q_2
\nonumber\\
&\quad+504\bigl(330q_3^{-2}+4224q_3^{-1}+7524+4224q_3+330q_3^{2}\bigr)(q_1^2q_2+q_1q_2^2)+\ldots,
\\
\chi_{10}&=-\tfrac14\Bigl[\bigl(q_3^{-1}-2+q_3\bigr)q_1q_2
-\bigl(2q_3^{-2}+16q_3^{-1}-36+16q_3+2q_3^{2}\bigr)(q_1^2q_2+q_1q_2^2)\Bigr]+\ldots,
\\
\chi_{12}&=\tfrac1{12}\Bigl[\bigl(q_3^{-1}+10+q_3\bigr)q_1q_2
+\bigl(10q_3^{-2}-88q_3^{-1}-132-88q_3+10q_3^{2}\bigr)(q_1^2q_2+q_1q_2^2)\Bigr]+\ldots,
\\
\chi_{5}&=\frac{\I}2\Bigl[\bigl(q_3^{-1/2}-q_3^{1/2}\bigr)(q_1q_2)^{1/2}
-\bigl(q_3^{-3/2}+9q_3^{-1/2}-9q_3^{1/2}-q_3^{3/2}\bigr)(q_1q_2)^{1/2}(q_1+q_2)\Bigr]+\ldots.
\end{align}
\end{subequations}
Several features are visible directly. The expansions of $E_4$ and $E_6$ are
symmetric under $q_3\to q_3^{-1}$, hence, even in $Z$, as required for even-weight
forms. Setting $q_3=1$ kills the leading coefficient of $\chi_{10}$ and reduces that
of $\chi_{12}$ to unity; more precisely,
\begin{equation}
  \chi_{10} = \sin^2(\pi Z)\,q_1q_2+\ldots\;\sim\;\pi^2Z^2q_1q_2\,,
\end{equation}
so $\chi_{10}$ has a double zero on $H_1$. Setting $q_3=1$ in the displayed
coefficients reproduces $E_4(T)E_4(U)$, $E_6(T)E_6(U)$ and $\Delta(T)\Delta(U)$ term
by term. For example, the $(q_1^2q_2+q_1q_2^2)$ coefficient of $\chi_{12}$ becomes
$(10-88-132-88+10)/12=-24$, the coefficient of $q^2$ in
$\Delta=q\prod(1-q^n)^{24}$. The half-integral powers in $\chi_5$ reflect its
character. Squaring reproduces $\chi_{10}$ order by order, while $\chi_5$ is odd in
$Z$ and has a simple zero on $H_1$.

%%%%%%%%%%%%%%%%%%%%%%%%%%%%%%%%%
\subsection{The \texorpdfstring{$\nu_\chi$}{nu-chi} transformation and the forms \texorpdfstring{$\chi_5,\chi_{30},\chi_{35}$}{chi5, chi30, chi35}}
\label{app:characters}

The possible one-dimensional characters are determined by the abelianization of
$\mathrm{Sp}(2g,\ZZ)$. For $g=1$ it is $\Z{12}$, so
$\SL{2,\ZZ}$ has twelve one-dimensional characters and integer powers of $\eta^2$
give familiar examples. For $g=2$ the abelianization is $\Z2$, giving the
trivial character and one non-trivial character $\nu_\chi(\gamma)=\pm1$. For
$g\ge3$ the group is perfect and has no non-trivial one-dimensional characters; this
statement should not be confused with half-integral-weight multiplier systems or
metaplectic covers.

The basic form carrying $\nu_\chi$ is $\chi_5$ of Appendix~\ref{app:theta}, with
\begin{equation}
    \chi_5(\gamma\Omega)=\nu_\chi(\gamma)\,j(\gamma,\Omega)^5\,\chi_5(\Omega)\,,
\end{equation}
so that $\chi_5^2=\chi_{10}$ has trivial character. This is an algebraic statement
about divisor multiplicity and by itself assigns no physical origin to $H_1$.

The fundamental odd-weight form with trivial character is the weight-$35$ cusp form,
defined geometrically as the Jacobian determinant of the four even generators,
\begin{equation}
    \chi_{35}(\Omega)\propto\det
    \begin{pmatrix}
        4E_4 & 6E_6 & 10\chi_{10} & 12\chi_{12} \\
        \partial_TE_4 & \partial_TE_6 & \partial_T\chi_{10} & \partial_T\chi_{12}\\
        \partial_UE_4 & \partial_UE_6 & \partial_U\chi_{10} & \partial_U\chi_{12}\\
        \partial_ZE_4 & \partial_ZE_6 & \partial_Z\chi_{10} & \partial_Z\chi_{12}
    \end{pmatrix},
    \label{eq:chi35-jacobian}
\end{equation}
where the first row follows from Euler's theorem for homogeneity in the modular
weights. Since $\chi_{10}=\chi_5^2$, the third column produces a factor of $\chi_5$, so
\begin{equation}
    \chi_{35}=\chi_5\,\chi_{30}\,,
\end{equation}
with $\chi_{30}$ carrying $\nu_\chi$, because $\chi_{35}$ has trivial character while
$\chi_5$ does not. Igusa's expansion of $\chi_{35}$ as an alternating sum over the
sixty azygous triples of even characteristics is homogeneous of degree $70$ in the
$\theta_m$, and since $\chi_5$ is the product of all ten even theta constants,
\begin{equation}
    \chi_{30}(\Omega)=\frac{\chi_{35}(\theta_m)}{\prod_{m\,\rm even}\theta_m(\Omega)}
\end{equation}
up to normalization, homogeneous of degree $60$. The Jacobian is antisymmetric under
exchange of the two diagonal moduli, so its square is symmetric and, by Igusa's
theorem, a weight-$70$ polynomial in the even generators:
\begin{equation}
    \chi_{35}^2=\chi_{10}\,P_{60}(E_4,E_6,\chi_{10},\chi_{12})\,,
    \qquad P_{60}=\chi_{30}^2 \,.
\end{equation}
Holomorphic forms with trivial character are built from $\Aev$ and
$\chi_{35}\Aev$, while those with the non-trivial character are built from
$\chi_5\Aev$ and $\chi_{30}\Aev$:
\begin{equation}
    M=\Aev\oplus\chi_{35}\Aev\,,
    \qquad
    M^{\nu_\chi}=\chi_5\Aev\oplus\chi_{30}\Aev \,,
    \label{eq:char-modules}
\end{equation}
the sums being direct because $\chi_5\Aev$ has odd and $\chi_{30}\Aev$ even weights.
Globally, forms in the non-trivial one-dimensional representation are not generated
by a single function. After restricting to weight zero and to a specified allowed
pole divisor, however, the parity arguments of \Cref{sec:chi10,sec:chi12,sec:chi35} reduce \Cref{eq:char-modules} to one, three
and one generators, respectively.

%%%%%%%%%%%%%%%%%%%%%%%%%%%%%%%%%
\subsection{Divisors of the generators}
\label{app:divisors}

\paragraph{$\chi_{10}$.}
From the product formula, $\chi_{10}$ vanishes exactly when at least one even theta
constant vanishes, which occurs exactly when $\Omega$ is modular equivalent to a
diagonal matrix. Its zero locus is therefore the divisor represented by $Z=0$, the
separating Humbert surface $H_1$. Under Torelli it is the decomposable
principally-polarized abelian-surface locus, reached from the smooth genus-two curve
locus by a separating stable degeneration. In the heterotic setting with one Wilson
line it is also the perturbative gauge-enhancement locus.

\paragraph{$\chi_{12}$.}
Let $\Amod=\Spg\backslash\Siegel$ and
$\mathcal D_{12}=\{[\Omega]\in\Amod\mid\chi_{12}(\Omega)=0\}$. This divisor should
not be confused with $H_1$: on the separating locus
$\chi_{12}(\diag(T,U))=\Delta(T)\Delta(U)$, and $\Delta$
has no zero in the upper half-plane, so $\mathcal D_{12}$ does not meet $H_1$ in the
interior of $\Amod$.

The interior divisor has an exact description in the invariant theory of binary
sextics \cite{Igusa:1960,Igusa:1962,Igusa:1967,CleryFaberVanderGeer:2017}. Every
point of $\Amod\setminus H_1$ is the principally polarized Jacobian of a smooth
genus-two curve
\begin{equation}
    C:\ y^2=F(x,1)\,,
    \qquad
    F(X,Y)=\sum_{r=0}^{6}\binom{6}{r}\mathfrak a_rX^{6-r}Y^r\,,
    \qquad \Disc(F)\neq0 \,.
\end{equation}
The classical quadratic invariant of the binary sextic is, up to a non-zero overall
normalization,
\begin{equation}
    \mathfrak a(F)=\mathfrak a_0\mathfrak a_6-6\mathfrak a_1\mathfrak a_5
    +15\mathfrak a_2\mathfrak a_4-10\mathfrak a_3^2 \,.
    \label{eq:sextic-invariant}
\end{equation}
Igusa's correspondence identifies the discriminant of the sextic with $\chi_{10}$
and the quadratic invariant, on the smooth-curve patch, with a non-zero multiple of
$\chi_{12}/\chi_{10}$. Hence, for a period matrix $\Omega_C$ of a smooth genus-two
curve,
\begin{equation}
    \chi_{12}(\Omega_C)=0\qquad\Longleftrightarrow\qquad
    \mathfrak a(F)=0 \,.
    \label{eq:chi12-characterization}
\end{equation}
Writing $t_2:\Mcurv\to\Amod$ for the Torelli map, the full interior zero divisor is
\begin{equation}
    \mathcal D_{12}=t_2\bigl(\{[C]\in\Mcurv\mid \mathfrak a(F_C)=0\}\bigr)\,,
\end{equation}
a complex codimension-one locus in the three-dimensional $\Amod$ rather than a
discrete set of points. The condition $\mathfrak a(F)=0$ is the
\emph{self-apolarity} condition for the binary sextic~\cite{DunajskiPenrose:2017}; it
does not describe a collision of branch points and, hence, does not represent a
degeneration of the curve. In this precise sense, $\mathcal D_{12}$ is a
well-defined modular divisor with no automatic massless-state interpretation.

\Cref{eq:chi12-characterization} also gives a direct construction of
numerical zeros. Choose $\mathfrak a_0,\ldots,\mathfrak a_5$ with
$\mathfrak a_0\neq0$ and set
$\mathfrak a_6=(6\mathfrak a_1\mathfrak a_5-15\mathfrak a_2\mathfrak a_4
+10\mathfrak a_3^2)/\mathfrak a_0$; retain only sextics with non-zero discriminant,
compute the normalized period matrix of the resulting curve and reduce it to a Siegel
fundamental domain. A distinguished point of $\mathcal D_{12}$ is $\OI$: its
stabilizer is cyclic of order ten, and after quotienting by $-\Id_4$ the
effective stabilizer has order five with $\det(C\OI+D)=\zeta$ for a suitable
generator \cite{Gottschling:1961,Ding:2020zxw,Leedom:2026nby}. Thus, any modular
form with no extra $\nu_\chi$ sign and weight $k$ obeys
$F_k(\OI)=\zeta^kF_k(\OI)$ and vanishes unless $5\mid k$. In particular,
$\chi_{12}(\OI)=0$.

For the present analysis we work directly with the standard absolute Igusa functions.
When three global modular coordinates are useful, we use these functions without
introducing a separate set of curve-invariant notation. In the normalization of
$E_4,E_6,\chi_{10},\chi_{12}$ adopted here, one conventional choice is
\begin{equation}
\begin{aligned}
 \mathfrak j_1&=2\,3^5\frac{\chi_{12}^5}{\chi_{10}^6}\,,\\
 \mathfrak j_2&=-2^{-3}3^3\frac{E_4\chi_{12}^3}{\chi_{10}^4}\,,\\
 \mathfrak j_3&=2^{-5}3\frac{E_6\chi_{12}^2}{\chi_{10}^3}
              +2^{-3}3^2\frac{E_4\chi_{12}^3}{\chi_{10}^4}\,.
\end{aligned}
\label{eq:absolute-Igusa}
\end{equation}
These three functions are a standard choice of absolute Igusa functions; the explicit
modular-form normalization above follows the conventional genus-two formulas used in
the computational literature \cite{BrokerLauter:2014}.\footnote{Conventions for binary-sextic invariants in the literature differ by overall signs and rescalings; throughout this paper $\mathfrak j_{1,2,3}$ are defined directly by \Cref{eq:absolute-Igusa}.}
They are algebraically independent on a generic patch and generate the genus-two modular function field rationally
\cite{Igusa:1960,Streng:2014}. Their precise overall normalizations are conventional
and play no role in the comparison with the basic ratios $\mathsf I_{\kappa}$ and
$\mathsf K_{\varsigma}$ of \Cref{sec:param}.

\paragraph{$\chi_{35}$ and $\chi_{30}$.}
The divisor of $\chi_{35}$ is reducible with two Humbert components modulo $\Spg$.
Because $\chi_{35}$ contains $\chi_5$ as a factor, it vanishes whenever
$\chi_{10}=0$, namely on $H_1$. In addition, $\chi_{35}$ is proportional to the
Jacobian determinant~\eqref{eq:chi35-jacobian} and is therefore antisymmetric under
$T\leftrightarrow U$. Since $E_4,E_6,\chi_{10},\chi_{12}$ are symmetric under that
exchange, which acts with $j_0=\det D=-1$ so that $j_0^k=1$ for even $k$, one has
$\partial_Tf|_{T=U}=\partial_Uf|_{T=U}$. The second and third rows of the Jacobian
then coincide, and the determinant vanishes identically on $T=U$. This is the
Humbert surface $H_4$, where a smooth abelian surface acquires an extra involution.
The divisor of $\chi_{30}$ contains $H_4$ but not $H_1$ as an irreducible component;
this does not imply that $\chi_{30}$ is non-zero at every point of $H_1$, since
intersections with other components may occur. %None of these facts implies any general incompatibility with the divisor of $\chi_{12}$.

%%%%%%%%%%%%%%%%%%%%%%%%%%%%%%%%%%%%%%%%%%%%%%%%%%%%%%%%%%%%%%%%%%%%%%%%%%%%%%%%
\section{Heterotic one-loop thresholds and gaugino condensation}
\label{app:thresholds}

This appendix collects the microscopic input summarized in \Cref{sec:setup}.

%%%%%%%%%%%%%%%%%%%%%%%%%%%%%%%%%%
\subsection{Wilsonian versus physical gauge couplings}

The holomorphic Wilsonian gauge kinetic function can be written schematically as
\begin{equation}
    f_a(S,\Omega)=S+\frac{1}{16\pi^2}\,\Delta_a^{\rm hol}(\Omega)\,,
\end{equation}
whereas the physical one-loop threshold is a real, non-holomorphic function obtained
from a modular integral~\cite{Kaplunovsky:1987rp,Dixon:1990pc}.
Only the holomorphic Wilsonian correction enters
the gaugino-condensate superpotential. Near a locus where states charged under $G_a$ become massless,
holomorphy of the Wilsonian coupling gives the familiar local logarithm
\begin{equation}
\label{eq:local-log-app}
    \Delta_a^{\rm hol}(\Omega)\simeq-2b_a^\star\log\Psi(\Omega)+\text{regular}\,,
\end{equation}
with $\Psi$ a local automorphic equation for the singular locus, while the physical
coupling contains the accompanying non-holomorphic terms required by target-space
anomaly cancellation. The full physical correction has the form
\begin{equation}
    \Delta_a^{\rm phys}(\Omega,\bar\Omega)
      =\int_{\mathcal F}\frac{\dd^2\tau}{\tau_2}
      \bigl[\mathcal B_a(\tau,\bar\tau;\Omega,\bar\Omega)-b_a\bigr]\,,
    \label{eq:DKL-integral}
\end{equation}
where $\mathcal B_a$ contains the Narain lattice sum with a gauge-charge insertion and
$\mathcal F$ is a worldsheet fundamental domain. The gauge factor therefore affects
more than an overall constant, but the allowed moduli dependence is controlled by the
same $\mathcal N=2$ BPS-saturated subsector. For one Wilson line the resulting
automorphic functions are genus-two Siegel modular forms or closely related lifts,
and the duality group is enlarged from $\mathrm O(2,2;\ZZ)$ to
$\mathrm O(2,3;\ZZ)\simeq\Spg$~\cite{Mayr:1995rx,Nilles:1997vk,Stieberger:1998yi,Leedom:2026nby}. The target
automorphic group is set by the lattice signature of the compactification, not by the
choice of hidden gauge group.

%%%%%%%%%%%%%%%%%%%%%%%%%%%%%%%%%%
\subsection{Threshold singularities and automorphic product formulas}
\label{app:holothresholds}

For thresholds that can be written as exponential or Borcherds lifts, the holomorphic
factor has a product representation
\begin{equation}
\label{eq:holofactor}
    \Psi(\Omega) = \e^{2\pi\I(\rho,\Omega)}
    \prod_{\ell>0}\bigl(1-\e^{2\pi\I(\ell,\Omega)}\bigr)^{c(\ell)}\,,
\end{equation}
where $\rho$ is the Weyl vector, and $\ell>0$
denotes lattice vectors in a chosen positive cone.
The exponents $c(\ell)$ are fixed by the Fourier
coefficients of the seed Jacobi form or elliptic genus. The polar data determine a divisor supported on rational quadratic, or
Humbert-type, loci, and these components have a direct interpretation in terms of BPS
states becoming massless~\cite{Borcherds:1995uml,Borcherds:1996uda,Harvey:1995fq}.

The form $\chi_{10}$ belongs to this class. In the $K3\times T^2$ example with one
Wilson line it is the exponential lift of the relevant weak Jacobi form and vanishes
on $H_1$, represented by $Z=0$, exactly where extra perturbative states can become
massless~\cite{Stieberger:1998yi,Leedom:2026nby}. The role of $\chi_{12}$ is
subtler. It appears explicitly in heterotic one-loop threshold formulae
\cite{Mayr:1995rx,Nilles:1997vk,Leedom:2026nby} and is non-zero at generic $Z=0$
because $\chi_{12}(\diag(T,U))\propto\Delta(T)\Delta(U)$, but its interior zero
divisor is not a Humbert enhancement divisor and is not known to arise from the same
elementary lift argument. Finally, $\chi_{35}$ appears in the universal threshold
structure of Nilles and Stieberger through factors such as
$[4\I\chi_{35}]^{1/2}$ together with powers of $\chi_{10}$~\cite{Nilles:1997vk}.
This establishes a top-down occurrence but does not imply that
an arbitrary isolated power of $\chi_{35}$ is generated in every compactification.

\subsection{Gaugino condensation and allowed automorphic powers}

For a pure hidden gauge sector, gaugino condensation generates
\begin{equation}
\label{eq:gaugino}
    W_{\rm gc}\propto\exp\Bigl(-\frac{8\pi^2}{c_a}f_a\Bigr)
\end{equation}
with $c_a$ the dual Coxeter number in the chosen normalization~\cite{Derendinger:1985kk,Dine:1985rz}.
If the holomorphic threshold contains \Cref{eq:local-log-app}, exponentiation gives a power of $\Psi$,
\begin{equation}
    W_{\rm gc}\propto\e^{-8\pi^2S/c_a}\,\Psi(\Omega)^{\,b_a^\star/c_a}\,,
\end{equation}
whose exponent is fixed by the spectrum and normalization conventions. Globally, this
must define a single-valued modular transformation law, which for the full genus-two
group may involve the non-trivial character, as for $\chi_5$ and $\chi_{30}$. Integer
and half-integer powers in explicit formulae should therefore be read as consequences
of the actual transformation law or character, not as a universal quantization theorem for $b_a^\star/c_a$.

%%%%%%%%%%%%%%%%%%%%%%%%%%%%%%%%%%%%%%%%%%%%%%%%%%%%%%%%%%%%%%%%%%%%%%%%%%%%%%%%%%%%%%%%%%%%
\section{Details of the \texorpdfstring{$H(\Omega)$}{H(Omega)} parametrizations}
\label{app:presentations}

This appendix supplies the counting behind \Cref{sec:param}, the explicit
order-by-order dictionary between the two forms of $H$, and the additional syzygies
needed when $H$ transforms with the extra $\nu_\chi$ sign.

\subsection{Derivation of the basic ratios}
\label{app:semigroup}

Because the even ring has only even weights, the weight-zero functions in Case (i) are
spanned by monomials $E_4^aE_6^b\chi_{12}^e/\chi_{10}^{\,r}$ whose exponents satisfy
\Cref{eq:semigroup-10}, and likewise for Case (ii) with
\Cref{eq:semigroup-12}. Adding exponent vectors adds weights, so the allowed
vectors close under addition and are generated by those that are not sums of two
allowed vectors. Enumerating them gives eight in the first case and six in the
second:
\begin{equation}
\begin{aligned}
  \chi_{10}:&\quad (a,b,e)=(1,1,0),(5,0,0),(2,0,1),(0,5,0),(0,3,1),(0,1,2),(1,0,3),(0,0,5)\,,\\
  \chi_{12}:&\quad (a,b,c)=(3,0,0),(0,2,0),(2,1,1),(1,0,2),(0,1,3),(0,0,6)\,,
\end{aligned}
\end{equation}
which are \Cref{eq:Igens,eq:Kgens} with $r=(4a+6b+12e)/10$ and $r=(4a+6b+10c)/12$. Two consequences are used in the main
text. First, indecomposability means precisely that a ratio is not a product of the
others, which makes the coordinate count at $\OI$ and $\OVI$ in
\Cref{sec:threeinvariants} equal to eight and six; the grading by pole order
$r$ makes the count unambiguous because all generators have $r\ge1$.
Second, the same congruence check shows that
$\widetilde P_{60}=P_{60}/\chi_{10}^6$ is a
polynomial in the basic ratios $\mathsf I_\kappa$.
Consequently, so are its powers, as required
for Case (iii).

\subsection{Complete syzygies}
\label{app:syzygies}

The polynomial relations among the basic ratios form toric syzygy ideals.
Their complete minimal generating sets are
\begin{equation}
\label{eq:Isyz}
\begin{aligned}
&\mathsf I_1^5-\mathsf I_2\mathsf I_4\,,\quad&
&\mathsf I_3^3-\mathsf I_2\mathsf I_7\,,\quad&
&\mathsf I_7^2-\mathsf I_3\mathsf I_8\,,\quad&
&\mathsf I_5^2-\mathsf I_4\mathsf I_6\,,\quad&
&\mathsf I_6^3-\mathsf I_5\mathsf I_8\,,\\
&\mathsf I_1\mathsf I_7-\mathsf I_3\mathsf I_6\,,\quad&
&\mathsf I_1^2\mathsf I_6-\mathsf I_3\mathsf I_5\,,\quad&
&\mathsf I_1\mathsf I_8-\mathsf I_6\mathsf I_7\,,\quad&
&\mathsf I_1^2\mathsf I_5-\mathsf I_3\mathsf I_4\,,\quad&
&\mathsf I_2\mathsf I_5-\mathsf I_1^3\mathsf I_3\,,\\
&\mathsf I_5\mathsf I_7-\mathsf I_1\mathsf I_6^2\,,\quad&
&\mathsf I_2\mathsf I_6-\mathsf I_1\mathsf I_3^2\,,\quad&
&\mathsf I_4\mathsf I_7-\mathsf I_1\mathsf I_5\mathsf I_6\,,\quad&
&\mathsf I_5\mathsf I_6^2-\mathsf I_4\mathsf I_8\,,\quad&
&\mathsf I_3^2\mathsf I_7-\mathsf I_2\mathsf I_8\,,
\end{aligned}
\end{equation}
and
\begin{equation}
\label{eq:Jsyz}
\begin{aligned}
&\mathsf K_3^2-\mathsf K_1\mathsf K_2\mathsf K_4\,,\quad &
&\mathsf K_4^3-\mathsf K_1\mathsf K_6\,,\quad&
&\mathsf K_5^2-\mathsf K_2\mathsf K_6\,,\\
&\mathsf K_3\mathsf K_4-\mathsf K_1\mathsf K_5\,,\quad &
&\mathsf K_3\mathsf K_5-\mathsf K_2\mathsf K_4^2\,,\quad&
&\mathsf K_4^2\mathsf K_5-\mathsf K_3\mathsf K_6 \,.
\end{aligned}
\end{equation}
Completeness can be checked systematically by organizing monomials according to their
exponent vectors. At a fixed multidegree, collect all products of basic ratios that
reproduce it and join two products whenever they share a common factor. The number
of independent new relations needed at that multidegree is the number of connected
components minus one. This gives exactly fifteen generating syzygies in Case (i),
\Cref{eq:Isyz}, and six in Case (ii), \Cref{eq:Jsyz}; no further relation is
needed beyond the degrees shown, and deleting any one of them leaves a set that no
longer implies it. Each relation is also verified directly by substituting the
definitions. For orientation, the multidegrees of the fifteen $\chi_{10}$ relations,
in the variables $(a,b,e)$ and in the order shown in \Cref{eq:Isyz}, are
\begin{equation}
\begin{aligned}
&(5,5,0),\ (6,0,3),\ (2,0,6),\ (0,6,2),\ (0,3,6),\\
&(2,1,3),\ (2,3,2),\ (1,1,5),\ (2,5,1),\ (5,3,1),\\
&(1,3,4),\ (5,1,2),\ (1,5,3),\ (0,5,5),\ (5,0,5).
\end{aligned}
\end{equation}
The last three,
$\mathsf I_4\mathsf I_7-\mathsf I_1\mathsf I_5\mathsf I_6$,
$\mathsf I_5\mathsf I_6^2-\mathsf I_4\mathsf I_8$ and
$\mathsf I_3^2\mathsf I_7-\mathsf I_2\mathsf I_8$, are the relations most easily
missed by inspection, and none follows from the first twelve. Two further relations
independent of the first ten are
$\mathsf I_5\mathsf I_7-\mathsf I_1\mathsf I_6^2$ and
$\mathsf I_2\mathsf I_6-\mathsf I_1\mathsf I_3^2$: the first ten imply only
$\mathsf I_1^2(\mathsf I_2\mathsf I_6-\mathsf I_1\mathsf I_3^2)\in\Isyz$, and one
cannot divide by $\mathsf I_1^2$.

%%%%%%%%%%%%%%%%%%%%%%%%%%%%%%%%%%%%%%%%%%%
\subsection{Pole-order expansion versus the basic-ratio parametrization}
\label{app:dictionary}

At every finite pole order the expansion~\eqref{eq:pole-order-expansion} is a
polynomial in the basic ratios; one simply decomposes each isobaric numerator into
monomials. For Case (i) the allowed weight-$10r$ monomials in $E_4,E_6,\chi_{12}$ are
\begin{align}
r=1:&\quad E_4E_6
&&\Longrightarrow\quad \frac{P_{10}}{\chi_{10}}=c_1\mathsf I_1\,,
\nonumber\\
r=2:&\quad E_4^5,\ E_4^2\chi_{12},\ E_4^2E_6^2
&&\Longrightarrow\quad \frac{P_{20}}{\chi_{10}^2}=c_2\mathsf I_2+c_3\mathsf I_3+c_9\mathsf I_1^2\,,\\
r=3:&\quad E_6^5,\ E_6^3\chi_{12},\ E_6\chi_{12}^2\,,\nonumber\\
&\quad\ \ E_4^3E_6\chi_{12},\ E_4^6E_6,\ E_4^3E_6^3
&&\Longrightarrow\quad \frac{P_{30}}{\chi_{10}^3}
=c_4\mathsf I_4+c_5\mathsf I_5+c_6\mathsf I_6+c_7\mathsf I_1\mathsf I_3+c_8\mathsf I_1\mathsf I_2+c_{10}\mathsf I_1^3,\nonumber
\end{align}
where at order three the terms with $c_7,c_8,c_{10}$ factorize into products of lower ratios.
For Case (ii),
\begin{align}
r=1:&\quad E_4^3,\ E_6^2 &&\Longrightarrow\quad d_1\mathsf K_1+d_2\mathsf K_2\,,\\
r=2:&\quad E_4^6,\ E_6^4,\ E_4^3E_6^2,\ E_4^2E_6\chi_{10},\ E_4\chi_{10}^2
&&\Longrightarrow\quad d_3\mathsf K_1^2+d_4\mathsf K_2^2+d_5\mathsf K_1\mathsf K_2+d_6\mathsf K_3+d_7\mathsf K_4\,,\nonumber
\end{align}
while $\mathsf K_5$ first appears at $r=3$ through $E_6\chi_{10}^3$, and the last
new ratio $\mathsf K_6$ only at $r=5$. The pole order at which a new ratio first
appears therefore keeps growing, which is why truncating at low pole order is not the
same as truncating the list of ratios.

\subsection{Useful adapted rational coordinates}
\label{app:adapted}

Only three weight-zero functions can be algebraically independent. A convenient
triple of algebraically independent rational coordinates adapted to the $\chi_{10}$ pole class is
\begin{equation}
    x_1=\mathsf I_1=\frac{E_4E_6}{\chi_{10}}\,,\qquad
    x_2=\mathsf I_2=\frac{E_4^5}{\chi_{10}^2}\,,\qquad
    x_3=\mathsf I_3=\frac{E_4^2\chi_{12}}{\chi_{10}^2}\,,
\end{equation}
in terms of which the remaining ratios are rational,
\begin{equation}
    \mathsf I_4=\frac{x_1^5}{x_2}\,,\qquad
    \mathsf I_5=\frac{x_1^3x_3}{x_2}\,,\qquad
    \mathsf I_6=\frac{x_1x_3^2}{x_2}\,,\qquad
    \mathsf I_7=\frac{x_3^3}{x_2},\qquad
    \mathsf I_8=\frac{x_3^5}{x_2^2}\,,
\end{equation}
and likewise
\begin{equation}
\begin{aligned}
    \mathsf K_1&=\frac{x_2}{x_3}\,,\quad &
    \mathsf K_2&=\frac{x_1^2}{x_3}\,, \quad&
    \mathsf K_3&=\frac{x_1x_2}{x_3^2}\,,\\
    \mathsf K_4&=\frac{x_2}{x_3^2}\,,\quad &
    \mathsf K_5&=\frac{x_1x_2}{x_3^3}\,, \quad&
    \mathsf K_6&=\frac{x_2^2}{x_3^5}\,.
\end{aligned}
\label{eq:K-in-x}
\end{equation}
These maps are used in Appendix~\ref{app:genvalues} as a cross-check on the reconstructed
values. They also make the coordinate artifact of \Cref{sec:threeinvariants} visible:
$\mathsf I_4=x_1^5/x_2$ looks singular where $x_2\propto E_4^5$ vanishes,
whereas $\mathsf I_4=E_6^5/\chi_{10}^3$ is manifestly regular there. Using the basic
ratios as generators avoids such artifacts, at the price of an overcomplete list
obeying the relations of Appendix~\ref{app:syzygies}.

\subsection{Additional syzygies for functions with the extra \texorpdfstring{$\nu_\chi$}{nu-chi} transformation}
\label{app:char-relations}

In Case (i), one generator is enough and the only relation is the closure
$\Ich^2=\widetilde P_{60}$ of \Cref{eq:Ichi-closure}. In Case (iii), the same holds
with $\Ich^{\pm1}$.

In Case (ii), three module generators are needed. Their module syzygies are non-trivial, so all three are required to parametrize the allowed functions. Setting
$\widehat P_{60}:= P_{60}/\chi_{12}^5=\chi_{30}^2/\chi_{12}^5$, which is itself a
Case-(ii) function, the linear relations among $L_1,L_2,L_3$ are
\begin{equation}
\begin{aligned}
 \mathsf K_3L_1&=\mathsf K_2L_2\,,\qquad&
 \mathsf K_6L_1&=\mathsf K_5L_3\,,\qquad&
 \mathsf K_4L_2&=\mathsf K_1L_3\,,\\
 \mathsf K_5L_1&=\mathsf K_2L_3\,,&
 \mathsf K_4^2L_1&=\mathsf K_5L_2\,,&
 \mathsf K_4^2L_1&=\mathsf K_3L_3\,,\\
 \mathsf K_1\mathsf K_4L_1&=\mathsf K_3L_2\,,&
 \mathsf K_6L_2&=\mathsf K_4^2L_3\,,
\end{aligned}
\label{eq:L-linear-complete}
\end{equation}
which follow by matching the exponents of $E_4,E_6,\chi_{10}$ in \Cref{eq:Lgens} modulo \Cref{eq:Jsyz}, and the products close according to
\begin{equation}
\begin{aligned}
    L_1^2&=\mathsf K_2\,\widehat P_{60}\,,\qquad &
    L_1L_2&=\mathsf K_3\,\widehat P_{60}\,,\qquad &
    L_1L_3&=\mathsf K_5\,\widehat P_{60}\,,\\
    L_2^2&=\mathsf K_1\mathsf K_4\,\widehat P_{60}\,, &
    L_2L_3&=\mathsf K_4^2\,\widehat P_{60}\,, &
    L_3^2&=\mathsf K_6\,\widehat P_{60}\,,
\end{aligned}
\end{equation}
each following directly from \Cref{eq:Lgens} and $\chi_{30}^2=P_{60}$.
Together with \Cref{eq:Jsyz}, these relations give the complete description of
functions obeying $H(\gamma\Omega)=\nu_\chi(\gamma)H(\Omega)$.

%%%%%%%%%%%%%%%%%%%%%%%%%%%%%%%%%%%%%%%%%%%%%%%%%%%%%%%%%%%%%%%%%%%%%%%%%%%%%%%%%%%%%%%%%%%%%%%%%%%%%%%%
\section{Fixed points and residual-symmetry analysis}
\label{app:fixedloci}

Following Gottschling's classification of the fixed loci of $\Spg$ in $\Siegel$~\cite{Gottschling:1961}
and its applications in Refs.~\cite{Ding:2020zxw,Nilles:2021glx} (summarized in \Cref{tab:fixedall}),
we now discuss the tangent data behind \Cref{tab:fixed0}. Let
$\gamma_0\in\Stab(\Omega_\star)$ and $M=C\Omega_\star+D$, so that
$j_0:= j(\gamma_0,\Omega_\star)=\det M$ and the tangent action is
$\delta\Omega\mapsto M^{-\top}\delta\Omega M^{-1}$. For a block-diagonal element built
from $\gamma_T\x\gamma_U\subset\SL{2,\ZZ}^2$ with
$M_T=c_TT+d_T$ and $M_U=c_UU+d_U$ one has $j_0=M_TM_U$ and diagonal tangent eigenvalues
\begin{equation}
\label{eq:tangent-eigen}
    (t_T,t_U,t_Z)=\bigl(M_T^{-2},\,M_U^{-2},\,(M_TM_U)^{-1}\bigr)\,,
\end{equation}
while the mirror element $T\leftrightarrow U$ has $C=0$ and $D$ a permutation, so
$j_0=\det D=-1$ and the tangent map exchanges $\delta T$ with $\delta U$ and fixes
$\delta Z$. In an eigenbasis where $R(\gamma_0)$ is diagonal, the conditions for the
first and second derivatives of a weight-$w$ object to survive are
\begin{equation}
\label{eq:derivative-selection}
    t_i=j_0^{\,w}\quad(\text{first})\,,
    \qquad
    t_it_j=j_0^{\,w}\quad(\text{second})\,,
\end{equation}
imposed simultaneously for all stabilizer generators.

The explicit data are as follows. At $\OI$ the effective stabilizer is $\Z5$
with $j_0=\zeta$ and $(t_1,t_2,t_3)=(\zeta,\zeta^4,\zeta^2)$. At $\OIII$, a
generating set is $S_T$, $S_U$ and the mirror, with $j_0(S_T)=j_0(S_U)=\I$ and
tangent eigenvalues $(-1,1,-\I)$ and $(1,-1,-\I)$ in the basis $(\delta T,\delta U,\delta Z)$. At $\OIV$ the analogous
generators use the order-six element with $M=\e^{\I\pi/3}$, giving
$(\rho^4,1,\rho^5)$ and $(1,\rho^4,\rho^5)$ with $\rho=\e^{\I\pi/3}$ and
$j_0=\rho$. At $\OVI$ the group is $\Z{12}$, generated by an element with
$M_T=\e^{\I\pi/3}$ and $M_U=\I$, so with $\xi=\e^{\I\pi/6}$ one has
$j_0=\xi^5$ and $(t_1,t_2,t_3)=(\xi^8,\xi^6,\xi^7)$. At $\OII$ and $\OV$ the
tangent representations are those in \Cref{eq:tangent-S4,eq:tangent-D12},
and $j_0$ is the sign character of $S_4$ and the $\Z2$ sign of $S_3\x\Z2$, respectively, as follows from $\det R=j_0^{-3}$.

At weight zero the first condition in \Cref{eq:derivative-selection} has no
solution at any of the six orbits, which gives \Cref{eq:auto-criticality}. For
the second condition, at $\OI$ the only pair with $t_it_j=1$ is $(1,2)$, so
$\mathcal Q_\star$ is spanned by one off-diagonal matrix of rank two. At $\OIII$ the
surviving invariant is $\delta T^2+\delta U^2$, of rank two; at $\OVI$ it is
$\delta U^2$, of rank one; and at $\OIV$ there is none. For $\OII$ and $\OV$ the
counts follow from the $\Sym^2$ decompositions in \Cref{sec:criteria}. These
are the entries of \Cref{tab:fixed0}.

Finally, \Cref{eq:derivative-selection} is valid at arbitrary weight, so only the
residue of $w$ modulo $d_\star$ can matter. We use this once for the holomorphic
matrix $\nabla_iD_jW$ defined in \Cref{eq:M-Wnonzero}: it carries $w=-35$ at
$\OI$, where $d_\star=5$, so $j_0^{\,w}=\zeta^{-35}=1$ and both conditions reduce to
their weight-zero form, giving $r_\star=2$. A systematic sweep over all weight
classes is possible but is not needed for the weight-zero $H$ imposed by the frame of
\Cref{sec:setup}.

\begin{table}[t!]
\centering
\small
\begin{tabular}{lllll}
\toprule
orbit & \multicolumn{4}{l}{values of the basic ratios}\\
\midrule
$\OI$ ($\Z5$) & \multicolumn{4}{p{0.78\textwidth}}{$\mathsf I_1=\cdots=\mathsf I_8=0$; outside Case-(ii) coordinate domain because $\chi_{12}=0$}\\
\midrule
$\OII$ ($S_4$)
 & $\mathsf I_1=-358400$ & $\mathsf I_2=-13107200000$ & $\mathsf I_3=\tfrac{128000}{3}$ & $\mathsf I_4=\mathsf I_1^5/\mathsf I_2$\\
 & $\mathsf I_5=\tfrac{449576960000}{3}$ & $\mathsf I_6=\tfrac{448000}{9}$ & $\mathsf I_7=-\tfrac{160000}{27}$ & $\mathsf I_8=\tfrac{200000}{243}$\\
 & $\mathsf K_1=-307200$ & $\mathsf K_2=3010560$ & $\mathsf K_3=2580480$ & $\mathsf K_4=-\tfrac{36}{5}$\\
 & $\mathsf K_5=\tfrac{1512}{25}$ & $\mathsf K_6=\tfrac{243}{200000}$ & &\\
\midrule
$\OV$ ($D_{12}$)
 & $\mathsf I_1=-518400$ & $\mathsf I_2=1343692800000$ & $\mathsf I_3=576000$ & $\mathsf I_4=\mathsf I_1^5/\mathsf I_2$\\
 & $\mathsf I_5=-59719680000$ & $\mathsf I_6=-128000$ & $\mathsf I_7=\tfrac{1280000}{9}$ & $\mathsf I_8=\tfrac{25600000}{729}$\\
 & $\mathsf K_1=2332800$ & $\mathsf K_2=466560$ & $\mathsf K_3=-2099520$ & $\mathsf K_4=\tfrac{81}{20}$\\
 & $\mathsf K_5=-\tfrac{729}{200}$ & $\mathsf K_6=\tfrac{729}{25600000}$ & &\\
\midrule
$\OIII$ & \multicolumn{4}{p{0.78\textwidth}}{outside Case-(i) coordinate domain because $\chi_{10}=0$; $\mathsf K_1=\mathfrak j(\I)^2=1728^2=2985984$ and $\mathsf K_2=\cdots=\mathsf K_6=0$}\\
$\OIV$ & \multicolumn{4}{p{0.78\textwidth}}{outside Case-(i) coordinate domain because $\chi_{10}=0$; $\mathsf K_2=(\mathfrak j(\omega)-1728)^2=1728^2$ and the other $\mathsf K_{\varsigma}$ vanish}\\
$\OVI$ & \multicolumn{4}{p{0.78\textwidth}}{outside Case-(i) coordinate domain because $\chi_{10}=0$; $\mathsf K_1=\cdots=\mathsf K_6=0$}\\
\bottomrule
\end{tabular}
\caption{Values of the basic ratios at the six
zero-dimensional elliptic fixed points.
The vanishing patterns follow from the fixed-point selection
rules in \Cref{sec:elliptic-conditions}.
The values of $\mathsf I_4$ at $\OII$ and $\OV$
are given by $\mathsf I_4=\mathsf I_1^5/\mathsf I_2$
from \Cref{eq:Isyz}.}
\label{tab:gens0}
\end{table}

%%%%%%%%%%%%%%%%%%%%%%%%%%%%%%%%%%
\subsection*{Basic ratios at elliptic fixed points}
\label{app:genvalues}

At the $S_4$ and $D_{12}$ orbits, the supersymmetric Minkowski construction imposes a
single linear condition $H(\Omega_\star)=0$ on the allowed coefficient space.
The corresponding elliptic points are \ac{CM}
points\footnote{\ac{CM} points correspond to Abelian surfaces
with a particularly rich set of algebraic self-maps.}, so their absolute Igusa invariants are
algebraic~\cite{Streng:2014},  though not necessarily
rational. For these points, high-precision theta
evaluation and rational reconstruction yield the
rational values listed in
\Cref{tab:gens0}. These values
satisfy the rational relations in
\Cref{eq:K-in-x} and the syzygies in
\Cref{eq:Isyz,eq:Jsyz}.
The exact values provide a direct arithmetic check of the fixed-point vacuum
conditions in \Cref{sec:vacua}.

{\small
%\bibliographystyle{OurBibTeX.bst}
%\bibliography{biblio}

\providecommand{\bysame}{\leavevmode\hbox to3em{\hrulefill}\thinspace}

}

\end{document}